\documentclass[journal,10pt,final,twoside]{IEEEtran}

\usepackage{tablefootnote,amsmath,mathtools,amssymb,cite,url,bm,graphicx,braket,float,blkarray}
\usepackage{multirow}
\usepackage[english]{babel}
\usepackage[utf8]{inputenc}
\usepackage[hidelinks,colorlinks=true,citecolor=black,linkcolor=black]{hyperref}

\newcommand{\Eb}{\mathbf{E}}
\newcommand{\Hb}{\mathbf{H}}
\newcommand{\Bb}{\mathbf{B}}
\newcommand{\Tc}{\mathcal{T}}
\newcommand{\Sh}{\hat{S}}
\newcommand{\ab}{\mathbf{a}}
\newcommand{\bb}{\mathbf{b}}
\newcommand{\epsh}{\hat{\varepsilon}}
\newcommand{\muh}{\hat{\mu}}
\newcommand{\rb}{\mathbf{r}}

\begin{document}
\bstctlcite{BSTcontrol}

\title{Integrated~Nonreciprocal~Photonic~Devices with~Dynamic~Modulation}

\author{Ian~A.~D.~Williamson, %
        Momchil~Minkov, %
        Avik~Dutt, %
        Jiahui~Wang, %
        Alex~Y.~Song, %
        and~Shanhui~Fan, %
\thanks{This work was supported in part by a MURI project from the U.S. Air Force Office of Scientific Research (AFOSR) (Grant No. FA9550-18-1-0379). (Corresponding author: Shanhui Fan.)} %
\thanks{Ian A. D. Williamson, Momchil Minkov, Avik Dutt, Alex Y. Song, and Shanhui Fan are with the Department of Electrical Engineering, Stanford University, Stanford, CA 94305 USA (e-mail: ian.williamson@ieee.org; mminkov@stanford.edu; avikdutt@stanford.edu; alexys@stanford.edu; shanhui@stanford.edu).} %
\thanks{Jiahui Wang is with the Department of Applied Physics, Stanford University, Stanford, CA 94305 USA (e-mail: jiahuiw@stanford.edu).}}

\markboth{Williamson \MakeLowercase{\textit{et al.}}: Integrated Nonreciprocal Photonic Devices with Dynamic Modulation}{Williamson \MakeLowercase{\textit{et al.}}: Integrated Nonreciprocal Photonic Devices with Dynamic Modulation}

\maketitle

\begin{abstract}
    Nonreciprocal components, such as isolators and circulators, are crucial components for photonic systems.
    In this article we review theoretical and experimental progress towards developing nonreciprocal photonic devices based on dynamic modulation. 
    In particular, we focus on approaches that operate at optical wavelengths and device architectures that have the potential for chip-scale integration. 
    We first discuss the requirements for constructing an isolator or circulator using dynamic modulation.
    We review a number of different isolator and circulator architectures, including waveguide and resonant devices, and describe their underlying operating principles. 
    We then compare these device architectures from a system-level performance perspective, considering how their figures of merit, such as footprint, bandwidth, isolation, and insertion loss scale with respect to device parameters.
\end{abstract}


\section{Introduction}

Integrated photonic platforms are driving a number of important technology advancements, including terabit-per-second optical communication links~\cite{marpaung_integrated_2013, marin-palomo_microresonator-based_2017, zhang_broadband_2019}, remote sensing for aerial radar~\cite{Fathpour2010, ghelfi_fully_2014, Serafino2019}, LIDAR phased arrays for self-driving vehicles~\cite{phare_silicon_2018, trocha_ultrafast_2018, suh_soliton_2018, sun_large-scale_2013}, quantum information processing~\cite{obrien_photonic_2009, wang_integrated_2019}, and even machine learning hardware accelerators~\cite{shen_deep_2017}.
Scaling up these technologies necessitates the development of optical circuits that combine thousands of elements, such as waveguides, switches, phase shifters, resonators, modulators, detectors, and sources.

Unlike integrated electronic devices, which are favorable for monolithic fabrication, photonic circuits typically require different materials for realizing high-performance active elements and low-loss passive elements.
Considerable progress has been made towards large-scale heterogeneous integration by addressing fundamental challenges in materials science and scaling up fabrication processes for high-performance photonic components from the lab to the foundry.
Examples include III-Vs with silicon \cite{komljenovic_photonic_2018, hiraki_heterogeneously_2017}, lithium niobate with oxide \cite{wang_integrated_2018a}, and diamond with silicon carbide and III-Vs \cite{Radulaski_diamond_2019}.
However, the integration of nonreciprocal elements, such as isolators and circulators, is still a major challenge because semiconductor materials conventionally used for photonic components are naturally reciprocal.
Typical isolator and circulator architectures rely on magneto-optical effects \cite{wang_optical_2005a, shoji_magnetooptical_2008, bi_onchip_2011, zhang_monolithic_2019}, which require the integration of yet another set of materials into an already highly complex fabrication flow.
More importantly, because magneto-optical effects tend to be fairly weak and the associated materials absorptive, magneto-optical isolators require careful management of the trade-off between the strength of the nonreciprocal response and signal attenuation.

The realization of compact and low-loss isolators and circulators would be a game changer for integrated photonics.
Isolators are two-port devices that allow light to propagate in one direction but absorb light that propagates in the opposite direction \cite{jalas_what_2013}.
These unidirectional elements play an important role in protecting sensitive laser sources from parasitic feedback, which results from both localized reflections off of individual optical circuit elements as well as distributed fabrication imperfections.
Isolators are crucial for preserving the stability of laser cavities as well as their spectral and noise properties \cite{tkach_regimes_1986}, making their integration a key step in scaling up coherent photonic integrated circuits.
For general purpose routing capabilities, optical circulators are crucial for their ability to route signals between more than two ports, based on a signal's propagation direction.
In optical communication networks, circulators allow transmitted and received signals to share a common physical channel \cite{sabharwal_inband_2014, zhang_selfinterference_2015, tang_fullduplex_2016, riaz_integration_2019}.
Moreover, by isolating high power transmitted signals from low-power received signals, the signal processing overhead required for interference cancellation can be significantly reduced \cite{sabharwal_inband_2014}.
In the quantum regime, circulators play a similarly important role in protecting sensitive readout circuits for qubits \cite{scheucher_quantum_2016, chapman_widely_2017, abdo_active_2019}.
Thus, the realization of low-loss and low-noise circulators will likely be a factor in scaling up quantum computing architectures to the point where they can out-perform classical processors \cite{bartlett_universal_2020}. 

Aside from fundamental issues of material integration, many important application spaces for integrated photonics are extremely sensitive to magnetic interference.
For example, optical readout from sensors based on atomic transitions requires careful control of the surrounding magnetic environment \cite{kitching_chipscale_2018}, meaning that alternatives to magneto-optics are required for nonreciprocal signal routing.
In recent years, the challenges outlined above have motivated the development of a different class of nonreciprocal components that are based on dynamic modulation, rather than magneto-optics.
Dynamically modulated components have long been used to break reciprocity in lower frequency electromagnetic regimes \cite{kamal_parametric_1960, baldwin_nonreciprocal_1961, maurer_lownoise_1963, anderson_reciprocity_1965a}, but promising advancements in optical modulator technologies means that high-performance nonreciprocal \textit{optical} components could soon be in reach.

In this article, we review theoretical and experimental progress on nonreciprocal devices based on dynamic modulation. 
While there is a large body of literature exploring these concepts in a variety of frequency ranges, here we focus on optical wavelengths and architectures with strong potential for chip-scale integration.
This review is organized as follows.
In Sec. \ref{sec:scattering} we begin by discussing the general requirements for achieving nonreciprocal responses in dynamically modulated optical elements, where we focus on a component-level scattering matrix perspective. 
In Sec. \ref{sec:foms} we define the figures of merit for nonreciprocal optical devices and then in Sec. \ref{sec:devices} we discuss the broad classes of device architectures based on the above requirements, and introduce their operating principles. 
In Sec. \ref{sec:mechanisms} we review conventional and emerging modulation mechanisms available for integrated optical devices.
In Sec. \ref{sec:comparison} we discuss the scaling of the device architectures from a system-level performance stand point, focusing on figures of merit such as physical footprint, modulation strength, operating bandwidth, and isolation contrast.
In Sec. \ref{sec:conclusion} we conclude with an outlook and discussion of the future prospects for dynamically modulated nonreciprocal devices.

\section{Scattering Matrix Perspective}
\label{sec:scattering}

\begin{figure}[tb]
    \centering
    \includegraphics{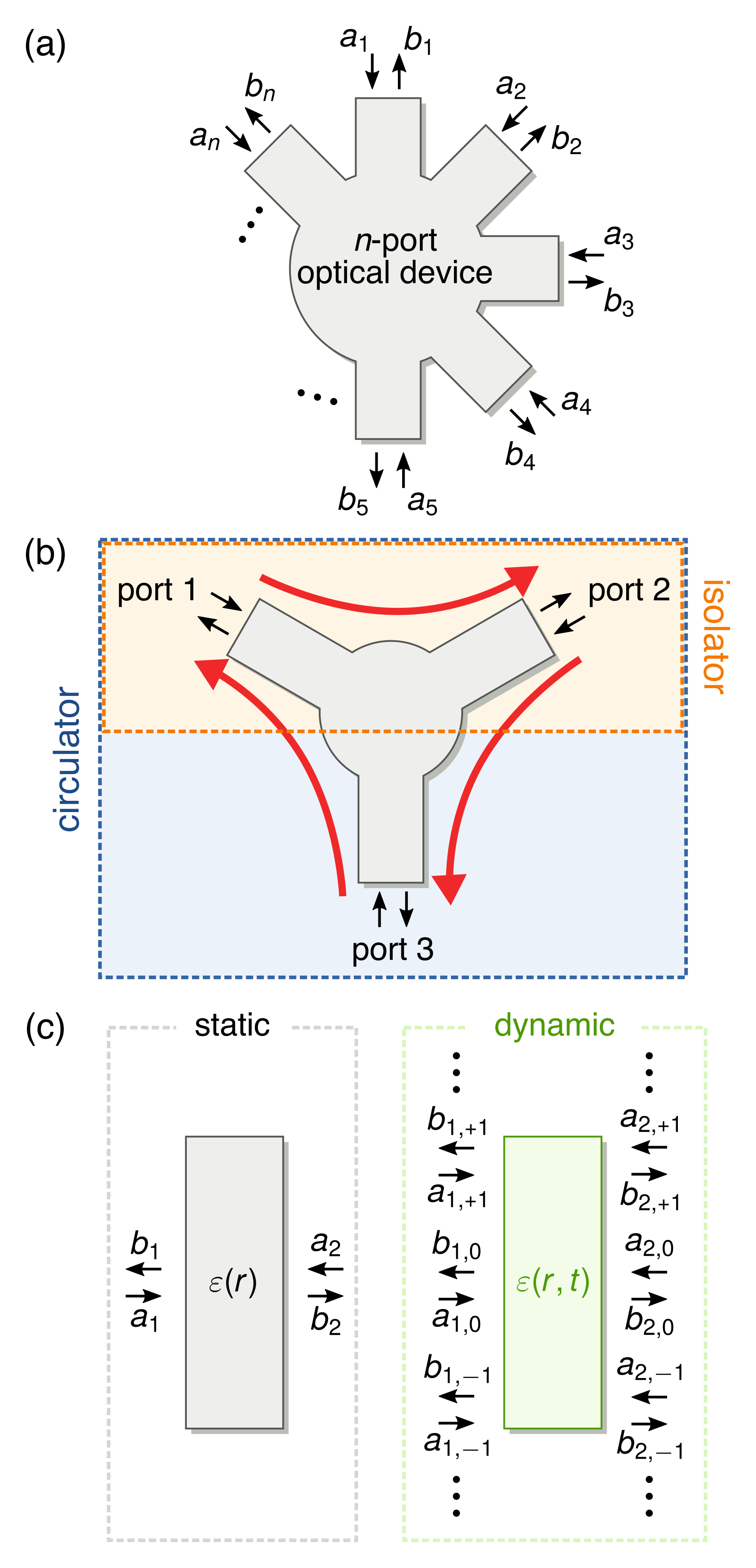}
    \caption{(a) Conceptual schematic of the scattering formalism for a general optical device with $n$ input-output ports. 
    The values of $a_i$ and $b_i$ represent the input and output amplitudes, respectively, of the port with index $i$. (b) 
    A two-port isolator is a sub-system of a three-port circulator. 
    The red arrows denote the non-zero elements of the circulator scattering matrix defined by (\ref{eqn:S_circulator}). 
    (c) Comparison of a static two-port device and a dynamically modulated two-port device. 
    The modulation extends the scattering response of each physical port to an infinite number of sidebands, or Floquet ports. 
    In practice, the number of Floquet ports can be truncated to a finite number based on the strength of the modulation.}
    \label{fig:scattering}
\end{figure}

We first review the scattering matrix formalism, which is a convenient mathematical description for the steady-state response of a multi-port optical device, shown schematically in Fig. \ref{fig:scattering}(a). 
For a linear and time-invariant device, the scattering matrix defines the steady-state input-output relations between different ports at a given angular frequency, $\omega$. 
Namely, 
\begin{equation}
    \begin{bmatrix}
    b_1 \\
    \vdots \\
    b_n
    \end{bmatrix} =
    \Sh(\omega) \begin{bmatrix}
    a_1 \\
    \vdots \\
    a_n
    \end{bmatrix},
    \label{eqn:Sdef}
\end{equation}
where $\ab = \begin{bmatrix} a_1 & \cdots & a_n \end{bmatrix}^T$ and $\bb = \begin{bmatrix} b_1 & \cdots & b_n \end{bmatrix}^T$ are, respectively, the incident and reflected amplitudes for $n$ different modes with harmonic time dependence $e^{j\omega t}$. 
These amplitudes are normalized such that the net power entering the device from port $i$ is $|a_i|^2 - |b_i|^2$. 
We emphasize that a \textit{port} corresponds to a mode of a physical channel.
In the case of a multi-mode waveguide each port corresponds to one of its modes. 
The scattering matrix is completely defined by the spatial distribution of the permittivity $\epsh$ and permeability $\muh$ tensors of the device. 

When a device has no gain or loss, both $\epsh$ and $\muh$ are Hermitian and power conservation dictates that $\Sh$ is unitary, meaning that
\begin{equation}
    \Sh \Sh^\dagger = \hat{I},
    \label{eqn:unitary}
\end{equation}
where $\hat{I}$ is the identity matrix. 
Reciprocity further constrains the form of the scattering matrix.
For a reciprocal device, if an excitation with a set of input amplitudes $\ab$ results in output $\bb$, then using $\bb$ as \textit{input} must produce $\ab$ at the outputs. 
This is true if and only if
\begin{equation}
    \Sh = \Sh^T.
    \label{eqn:reciprocal}
\end{equation}

Isolators and circulators are nonreciprocal devices, and must therefore violate (\ref{eqn:reciprocal}). 
The canonical scattering matrix for an ideal optical isolator is
\begin{equation}
    \Sh = \begin{bmatrix}
    S_{11} & S_{12} \\
    S_{21} & S_{22}
    \end{bmatrix} = \begin{bmatrix}
    0 & 0 \\
    1 & 0
    \end{bmatrix}, 
    \label{eqn:S_isolator}
\end{equation}
describing a two-port device in which power is fully transferred from port 1 to port 2, and fully attenuated when going in the opposite direction [Fig. \ref{fig:scattering}(b)].
However, this can only be achieved in a device with loss because the scattering matrix in (\ref{eqn:S_isolator}) does not satisfy the condition given by (\ref{eqn:unitary}). 
Specifically, in the two port device described by (\ref{eqn:S_isolator}), power injected into port 2 is lost. 
More generally, one can prove that for a two-port device, energy conservation alone requires that $\vert S_{12} \vert = \vert S_{21} \vert$. 
Thus, one cannot construct an energy conserving two-port isolator.
The simplest scattering matrix that is both \textit{nonreciprocal} and \textit{unitary} is that of a circulator, where
\begin{equation}
    \Sh = 
    \begin{bmatrix}
    0 & 0 & 1 \\
    1 & 0 & 0 \\
    0 & 1 & 0
    \end{bmatrix}.
    \label{eqn:S_circulator}
\end{equation}
This routing behavior is shown schematically in Fig. \ref{fig:scattering}(b).

It is useful for our discussion to interpret an isolator as a subsystem of a circulator, in which no power is allowed to enter from the third port. 
In other words, the scattering matrix defined by (\ref{eqn:S_isolator}) can be interpreted as the top left corner of the scattering matrix in (\ref{eqn:S_circulator}). 
Intuitively, this means that an isolator can dissipate the backward power entering from port 2 in a number of ways, including material absorption, routing to another output waveguide, scattering into the surrounding environment, or some linear combination of the above. 
All of these dissipation pathways can be in terms of extra ports, such that the full scattering matrix is unitary and lossless. 
Apart from accounting for the exact power balance of the device, an advantage of conceptualizing isolators in this way is that the relationship between nonreciprocity and time-reversal (TR) symmetry can be made explicit. 

The time-reversal operation $\Tc$ denotes how physical quantities change upon reversing the flow of time, e.g. $t \rightarrow -t$ \cite{haus_fields_1984}. 
When the properties of a material are invariant under time-reversal, e.g. $\Tc(\epsh) = \epsh$ and $\Tc(\muh) = \muh$, then any mode with electric and magnetic fields $\Eb$ and $\Hb$ at frequency $\omega$ has a time-reversed counterpart with fields $\Eb^*$ and $-\Hb^*$ at frequency $\omega^*$. 
From the perspective of the scattering process, the time-reversed counterpart to $(\ref{eqn:Sdef})$ is $\ab^* = \Sh \bb^*$. 
Thus, for a TR invariant device, $\ab = \Sh \Sh^*\ab^*$, or
\begin{equation}
    \Sh^* = \Sh^{-1}.
    \label{eqn:tr_S}
\end{equation}
The combination of (\ref{eqn:unitary}) and (\ref{eqn:tr_S}) implies (\ref{eqn:reciprocal}), and so a nonreciprocal, lossless device must necessarily break time-reversal symmetry. 
In order to construct a circulator (and hence an isolator), we must necessarily have either $\Tc(\epsh) \neq \epsh$ or $\Tc(\muh) \neq \muh$. 
Conventionally, optical devices have achieved this through magnetically biased gyrotropic materials \cite{shoji_magnetooptical_2008, bi_onchip_2011, zhang_monolithic_2019}, which have a permittivity tensor $\epsh$ that depends on the magnetic bias, $\Bb$.
Because magnetic fields flip sign under time-reversal, $\Tc(\Bb) = -\Bb$, we would then generally have $\Tc(\epsh(\Bb)) = \epsh(-\Bb) \neq \epsh(\Bb)$.

Alternatively, time-reversal symmetry can be broken in dynamically modulated devices, which is the central focus of this review.
Because we focus on devices operating in the optical regime, we always assume a non-magnetic material response ($\mu_r$ = 1) and, for simplicity, an isotropic scalar permittivity, $\varepsilon(\rb, t)$ that depends on both position and time.
When the modulation is \textit{periodic} with period $T$, a dynamic steady state does exist and oscillates in time with the same period, $T$.
In this case, the scattering response can be generalized into a \textit{Floquet} scattering matrix, assuming that only the interior region of the optical device varies in time, while the ports remain static.
In such a Floquet scattering framework, the Fourier expansion of the periodic function $a_i(t)$ allows the input amplitude in every port $i$ to be expanded as 
\begin{equation}
    a_i(t) e^{j\omega t} = \sum_p a_{i,p} e^{j(\omega +p\Omega)t}, \label{eq:a_expansion}
\end{equation}
where $\Omega = 2\pi/T$. 
A similar expansion can also be defined for the output amplitudes.
In practice, for a given modulation amplitude, there is negligible energy occupying sidebands above some threshold, meaning that the summation in (\ref{eq:a_expansion}) can be truncated to $|p| < P$ for some integer $P$.
Such a truncation allows us to define the incident and reflected amplitudes, respectively, as
\begin{align}
    \ab &= \begin{bmatrix} a_{1,-P} & \cdots & a_{1, P} & \cdots & a_{n, -P} & \cdots & a_{n, P} \end{bmatrix}^T \label{eqn:a_floquet} \\ 
    \bb &= \begin{bmatrix} b_{1,-P} & \cdots & b_{1, P} & \cdots & b_{n, -P} & \cdots & b_{n, P} \end{bmatrix}^T, 
\label{eqn:b_floquet}
\end{align}
which are linked by the finite-dimensional scattering matrix, $\bb = \Sh(\omega)\ab$ \cite{Moskalets2002, Tymchenko2017} relating the input and output through all \textit{sidebands} oscillating at $\omega+p\Omega$, for all ports [Fig. \ref{fig:scattering}(c)].
In this framework, we refer to a port as denoted by $i,\ p$ as a Floquet port. 
Here, $i$ indexes the physical port, and $p$ indexes the sideband. 
Additionally, the amplitudes are assumed to be normalized such that the net photon flux entering the device from a Floquet port $i, p$ is $(\omega + p\Omega)(|a_i|^2 - |b_i|^2)$, in order to maintain $\Sh \Sh^\dagger = I$ for a lossless device. 
In other words, while power is not necessarily conserved in the presence of dynamic modulation, the total photon number flux is conserved.
With this definition, the concept of the scattering matrix which is commonly defined for static devices is generalized to dynamically modulated devices.

\begin{figure}[tb]
    \centering
    \includegraphics{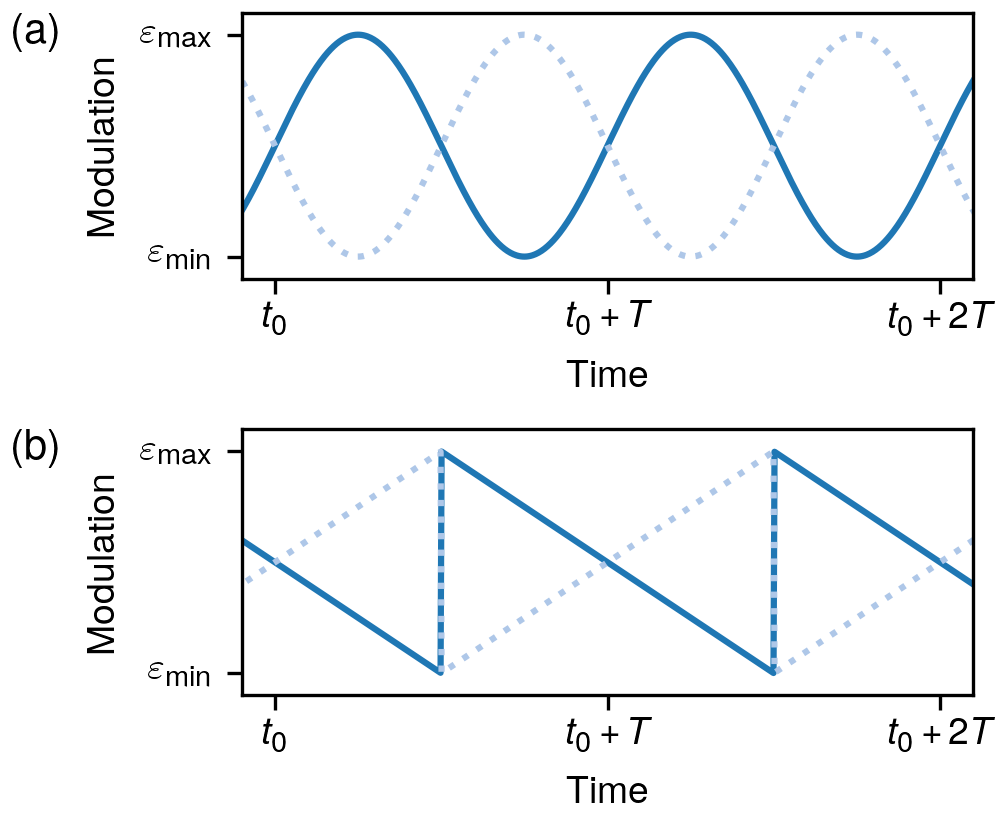}
    \caption{(a) A spatially uniform modulation with a sinusoidal time dependence does not break the generalized time-reversal symmetry defined by (\ref{eqn:tr_eps}), because the modulation (solid line) and its time-reversed version (dashed line) are identical up to an offset along the time axis. (b) A sawtooth wave is an example of modulation that breaks the generalized time-reversal symmetry defined by (\ref{eqn:tr_eps}), because the modulation (solid line) and its time-reversed version (dashed line) are not identical for \textit{any} offset along the time axis.}
    \label{fig:modulation}
\end{figure}

A nonreciprocal dynamically modulated device has an asymmetric scattering matrix because it does not satisfy the condition in (\ref{eqn:reciprocal}). 
In order to achieve this, the modulation must break time-reversal symmetry. 
Naively, one might expect that any modulation waveform with
\begin{equation}
    \varepsilon(\rb, t) \ne \varepsilon(\rb, -t)
    \label{eqn:tr_eps_naive} 
\end{equation}
will break time-reversal symmetry. 
However, here we are considering a steady state response, which is independent of the time origin.
Thus, to create a nonreciprocal dynamically modulated device, (\ref{eqn:tr_eps_naive}) must be satisfied independently of the time origin. 
To emphasize this point, we define a \textit{generalized time-reversal symmetry}: a time-modulated device is defined to maintain a generalized time-reversal symmetry if the condition
\begin{equation}
    \varepsilon(\rb, t - t_0) = \varepsilon(\rb, -t - t_0),
    \label{eqn:tr_eps} 
\end{equation}
is satisfied for at least one value of $t_0$. 
A nonreciprocal dynamically modulated device can only be constructed if the generalized time-reversal symmetry defined by (\ref{eqn:tr_eps}) is not satisfied for all choices of $t_0$. 
It has been noted in \cite{fang_photonic_2012, fang_realizing_2012} that a change of time origin $t_0$ corresponds to a gauge transformation of the photon wave function. 
Therefore, (\ref{eqn:tr_eps}) is essentially a gauge-invariant definition of time-reversal symmetry for the steady state response of a dynamically modulated device. 
This is illustrated in Fig. \ref{fig:modulation} for a spatially uniform modulation.
The sinusoidal time dependence shown in Fig. \ref{fig:modulation}(a) obeys the generalized time-reversal symmetry condition given by (\ref{eqn:tr_eps}), while the sawtooth modulation shown in Fig. \ref{fig:modulation}(b) does not. 
In Sec. \ref{sec:devices} we will describe several specific isolator designs in more detail that use dynamic modulation to violate the condition defined in (\ref{eqn:tr_eps}).
Additionally, we will highlight how many of these designs achieve isolation using signal loss pathways, as shown conceptually in Fig. \ref{fig:scattering}(b).

\section{Device Figures of Merit}
\label{sec:foms}

In this section we define the figures of merit for nonreciprocal devices that are relevant to their performance in integrated photonic circuits.
Here we discuss these figures of merit in terms of a two-port isolator, but as discussed in the previous section and illustrated in Fig. \ref{fig:scattering}(b), these definitions can be applied to any pair of ports in a circulator.

Throughout this review, we use the convention that the \textit{forward} direction of an an isolator refers to the direction in which it transmits signals, while the \textit{backward} direction refers to the direction in which it isolates signals, or suppresses their transmission.
Note that in some references this convention is reversed.

Some of the most important figures of merit and performance characteristics are defined as follows:
\begin{itemize}
    \item \textbf{Isolation ratio.}
    The isolation ratio, or often just referred to as the \textit{isolation}, refers to the ratio of the transmission coefficient in the forward direction to the transmission coefficient in the backward direction.
    An ideal isolator should provide infinite signal isolation, meaning that there is no signal transmission in the backward direction.
    \item \textbf{Bandwidth.}
    The bandwidth of an isolator generally refers to the frequency extent of a signal that it isolates, or suppresses, in the backward direction and that it transmits in the forward direction.
    In some devices, the forward and backward bandwidth may be different.
    For example, a device may provide isolation over a much narrower bandwidth in the backward direction than it provides for signal transmission in the forward direction.
    \item \textbf{Signal distortion.}
    Over its operating bandwidth, an ideal isolator should transmit a signal in the forward direction without distortion. 
    Examples of distortion can include changes to the pulse linewidth or chirping.
    An ideal isolator should have a linear phase response over its bandwidth with uniform signal transmission.
    In dynamically modulated isolators, an additional concern could be harmonic signal distortion from the modulation.
    \item \textbf{Insertion loss.}
    The insertion loss refers to the attenuation experienced by a signal propagating in the forward direction of the isolator.
    An ideal isolator should provide unity transmission, and thus zero insertion loss, in the forward direction.
    \item \textbf{Return loss.}
    The return loss, or reflection, refers to the portion of a signal returned to the input port of the isolator.
    Following the discussion on insertion loss above, an ideal isolator should have zero return loss for signals in the forward direction.
    This characteristic is especially important for laser protection because signals reflected into the laser cavity can result in instability and performance degradation.
    \item \textbf{Footprint.}
    The footprint of an integrated optical isolator could generally be defined as the physical area that it occupies on a chip. 
    In some devices, such as long optical waveguides, we interchangeably use footprint to refer to just the waveguide length, which is the dominant device dimension.
    \item \textbf{Tunability.}
    The tunability of an isolator refers to the extent to which the nonreciprocal response can be tuned via some parameter of the device, e.g. by reconfiguring the isolator at run time.
    For example, in certain isolator designs that we consider the spectral response can be shifted, allowing the isolator to operate in different frequency bands.
    \item \textbf{Power consumption.} 
    Unlike magneto-optical isolators, dynamically modulated isolators require an active modulation to enable their nonreciprocal response.
    Therefore, the power consumption associated with generating the modulation can be an important consideration.
    \item \textbf{Robustness.}
    The performance of an ideal isolator should be robust to environmental factors and fabrication variations.
    For example, an ideal isolator would have a response that is stable, or at least predictable, to changes in ambient temperature.
\end{itemize}

\section{Device Architectures}
\label{sec:devices}

We now discuss the device architectures of dynamically modulated isolators and circulators based on the requirements discussed in Sec. \ref{sec:scattering}.
In this section, we focus our analysis on introducing the operating principles for each device and then in Sec. \ref{sec:comparison} we provide a more comprehensive comparison of the different device architectures.
We organize our discussion here by classifying the devices into two broad categories: modulated optical waveguides and modulated optical resonators.

\subsection{Modulated Waveguides}
\label{sec:devices_waveguides}

\begin{figure*}[tb]
    \centering
    \includegraphics{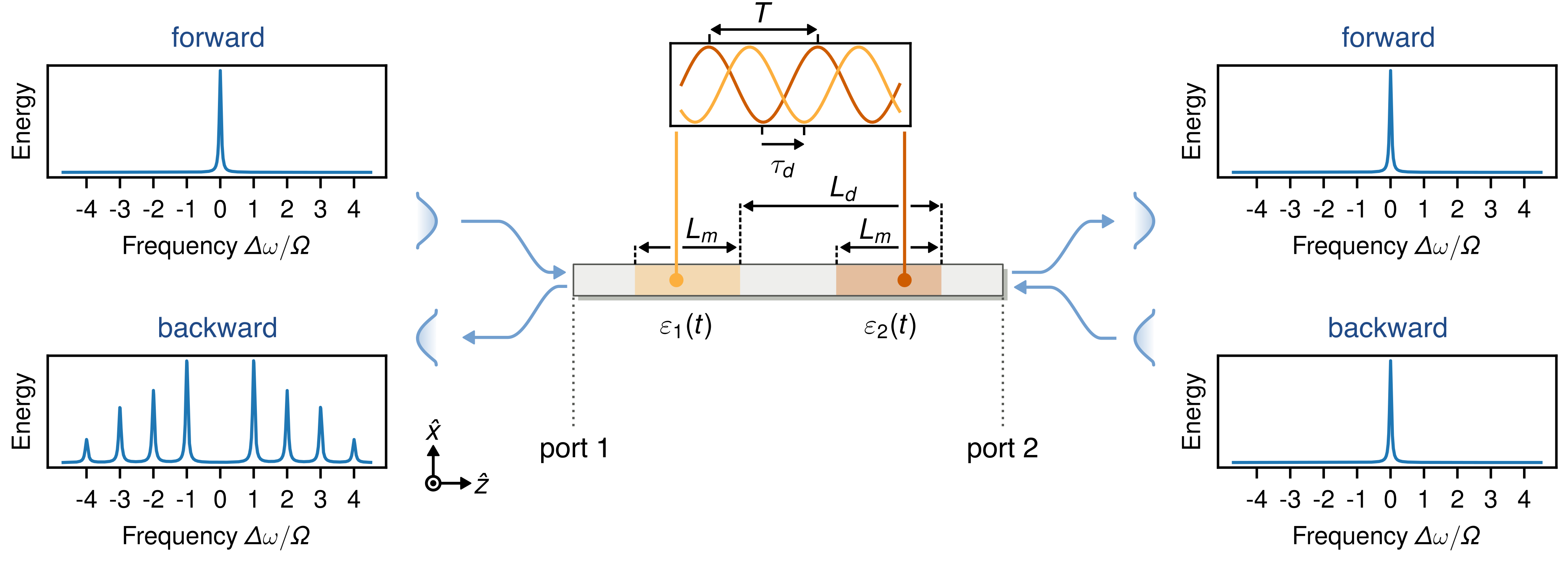}
    \caption{Optical isolator architecture based on tandem phase modulators, as proposed in \cite{doerr_optical_2011}. 
    Two waveguide modulators are separated by a passive section of waveguide that acts as a delay line. 
    The two modulators are modulated with a relative phase delay to create a nonreciprocal frequency-converter. 
    In the forward direction, signal energy remains at the input frequency, $\omega$. In the backward direction, signal energy is transferred completely to sidebands at intervals of the modulation frequency, $\Omega = 2\pi/T$.
    A passband filter, centered around the input signal frequency, is required to filter the sidebands in the backward direction and to allow the device to operate as an isolator.}
    \label{fig:waveguide_tandem}
\end{figure*}

Single-mode waveguides are devices that have only two physical ports (i.e. each end of the waveguide).
Constructing an optical isolator using such two-port devices necessitates that the scattering response be extended to additional ports, as shown in Fig. \ref{fig:scattering}(b). 
This can always be achieved using dynamic modulation which, as illustrated in Fig. \ref{fig:scattering}(c), couples signal energy between different Floquet ports.

\subsubsection{Tandem phase modulators}
\label{sec:tandem}

In this section, we review the architecture for a dynamically modulated isolator based on a sequence of two standing-wave modulators separated by a passive delay line.
This so-called tandem phase modulator isolator, as proposed in \cite{doerr_optical_2011}, is shown schematically in Fig. \ref{fig:waveguide_tandem}. 
Although such modulation locally satisfies the generalized time-reversal symmetry, defined in (\ref{eqn:tr_eps}), the device as a whole does break generalized time-reversal symmetry through a relative phase shift between the modulated regions.
Under a specific configuration that we describe below, this device functions as an asymmetric frequency converter between the forward and backward directions.

To review the concept of the device, we consider a model system consisting of a single-mode dielectric slab waveguide with relative permittivity, $\varepsilon_r$.
As shown in Fig. \ref{fig:waveguide_tandem}, optical signals propagate in the fundamental waveguide mode along the $z$-direction.
The permittivity of the two modulated waveguide segments have a time-dependence of
\begin{align}
	\frac{\varepsilon_1(t)}{\varepsilon_0} &= \varepsilon_r + \Delta{\varepsilon_r} \cos(\Omega t)
	\label{eq:epsr_tandem1} \\
	\frac{\varepsilon_2(t)}{\varepsilon_0} &= \varepsilon_r + \Delta{\varepsilon_r} \cos(\Omega t + \theta),
	\label{eq:epsr_tandem2}
\end{align}
where $\Delta \varepsilon_r / \varepsilon_r$ is defined as the modulation strength, $\Omega$ is the modulation frequency and $\theta$ is the phase difference between the two modulators.
The modulation is applied uniformly along the $x$-direction of the waveguide and each modulator has a length of $L_m$. 
Between the two modulator is a static waveguide of length $L_d-L_m$ that acts as a delay line. 

The device can be understood conceptually from the perspective of a time-dependent optical transmission.
Under the modulation defined by (\ref{eq:epsr_tandem1}) and (\ref{eq:epsr_tandem2}), a signal with input frequency $\omega$, will experience a time-dependent phase delay of $\phi_1(t) = A\cos(\Omega t)$ from passing through the first modulator, where the amplitude of the phase delay is $A = (\Delta\varepsilon_r/\varepsilon_r) L_m\omega/2v_g$.
We emphasize that $A$ is determined by both the modulator length, $L_m$ and the strength of the modulation, $\Delta\varepsilon_r / \varepsilon_r$. 
Similarly, the second modulator will introduce an additional time-dependent phase delay for the optical signal of $\phi_2(t) = A\cos (\Omega t + \theta)$, where $\theta$ is the relative phase difference of the second modulating wave with respect to the first.
The delay line results in a time delay $\tau_d=L_d/v_g$ for the optical signal propagating between the modulators, where $v_g$ is the group velocity of the optical signal.
The analysis above assumes that the waveguide is operating in an optical bandwidth with low dispersion, e.g. $\Delta \epsilon_r / \epsilon_r \approx 2\Delta n/ n_g \approx 2\Delta n/n$.

Therefore, in the \textit{forward} direction, an optical signal at $\omega$ entering port 1 has a time-dependent transmission of
\begin{equation}
    \label{eqn:forward_doerr}
    \begin{aligned}
    T_f(t) =& \exp[-j\phi_1(t-\tau_d)] \cdot \exp[ -j \phi_2(t)] \\
    =& \exp\hspace{-3pt}\left[-j 2A\cos\hspace{-2pt}\left(\Omega t + \frac{\theta - \Omega \tau_d}{2}\right) \cdot \cos\hspace{-2pt}\left(\frac{\theta + \Omega\tau_d}{2}\right)\right].
    \end{aligned}
\end{equation}
In the \textit{backward} direction, an optical signal at $\omega$ entering port 2 has a time-dependent transmission of
\begin{equation}
    \label{eqn:backward_doerr}
    \begin{aligned}
    T_b(t) =& 
    \exp[-j{\phi_2\left(t-\tau_d\right)}] \cdot \exp[-j\phi_1(t)] \\
    =& \exp\hspace{-3pt}\left[-j 2A\cos\hspace{-2pt}\left(\Omega t + \frac{\theta - \Omega \tau_d}{2}\right) \cdot \cos\hspace{-2pt}\left(\frac{\theta - \Omega \tau_d}{2}\right)\right] 
    \end{aligned}
\end{equation}
By designing the device to satisfy the condition $\theta = \pi/2$ and $\Omega \tau_d = \pi /2$~\cite{doerr_optical_2011}, the \textit{forward} direction has a transmission of $T_f(t) = 1$. 
The \textit{backward} has a transmission of $T_b(t) = \exp[-j2A\cos (\Omega t)] = \sum_n (-j)^n J_n (2A) \exp(jn \Omega t)$, where the last step utilizes the Jacobi-Anger expansion and $J_n(x)$ is an $n$-th order Bessel function of the first kind. 
Additionally, by designing the modulator length $L_m$ such that $J_0(2A) = 0$, the input signal energy in the \textit{backward} direction is completely shifted to the sideband frequency components.

From a Floquet point of view, in the \textit{forward} direction, an input signal with frequency $\omega$ is scattered into a number of Floquet modes at frequencies ($\omega \pm \Omega$, $\omega \pm 2\Omega$, $\omega \pm 3\Omega$, $\ldots$) by the first modulator. 
When passing through the second modulator, these Floquet modes are completely scattered back to the original signal at a frequency $\omega$. 
However, for the \textit{backward} direction, due to the nonreciprocal phase introduced by the time modulation, the energy of the input signal is shifted to other frequencies. 
Therefore to operate as an isolator, a band pass filter must be added such that the Floquet modes with frequency $\omega + p\Omega$ for $p\ne0$ are absorbed.
According to (\ref{eqn:forward_doerr}) and (\ref{eqn:backward_doerr}), the condition for achieving the strongest nonreciprocal response is that $\phi_1(t-\tau_d) + \phi_2(t) \neq \phi_2(t-\tau_d) + \phi_1(t)$.

From Fig. \ref{fig:waveguide_tandem}, we observe that the bandwidth of this design is limited by the modulation frequency, $\Omega$. 
One approach for extending the bandwidth to multiples of $\Omega$ is to utilize a configuration consisting of multiple parallel modulators that cancel additional sidebands, as proposed in \cite{doerr_silicon_2014}.
Another route towards an increased bandwidth is by utilizing non-sinusoidal phase modulation, e.g. with a square wave modulation \cite{galland_broadband_2013, yang_broadband_MZI_2014}.

We now discuss the total length $L = L_m + L_d$ of the tandem modulator isolator design. 
For the design where $\theta = \pi/2$, the modulation length required for each modulated segment is
\begin{equation}
    L_m = \frac{2 A_0 v_g}{(\Delta \varepsilon_r/\varepsilon_r) \omega},
    \label{eqn:tandem_Lm}
\end{equation}
which is on the order of few millimeters and does not depend on the modulation frequency. Here $2A_0\approx 0.77\pi$ such that $J_0(2A_0) = 0$. 
With a small modulation frequency on the order of several MHz, the delay-line length is
\begin{equation}
    L_d = \frac{\pi v_g}{2\Omega},
    \label{eqn:tandem_Ld}
\end{equation}
which can easily be on the order of several meters \cite{yang_broadband_MZI_2014}. 
Thus, the footprint of the original tandem phase modulator isolator is dominated by the length of the delay line, e.g. $L\approx L_d \sim 1/\Omega$.
The observation that $L_d \geq L_m$, leads to a constraint on the modulation strength of
\begin{equation}
    \frac{\Delta \varepsilon_r}{\varepsilon_r} \geq 1.54 \pi \frac{\Omega}{\omega},
    \label{eqn:tandem_constrain}
\end{equation}
which implies that the tandem phase modulator isolator should also be able to operate in a regime with stronger modulation. 
However, increasing the modulation strength alone is not sufficient to reduce the total length of the device, which is dominated by the length of the delay line given in (\ref{eqn:tandem_Ld}).
One modification to make the device more compact is to adjust the phase difference $\theta$ between two modulation segments such that the device needs a longer modulation length $L_m^\prime = L_m/\sin(\theta)$ but a shorter delay line length $L_d^\prime = 2 (1 - \theta/\pi) L_d$, as demonstrated in \cite{lin_compact_2019}. 
In this modified design, the total length can be approximated as
\begin{equation}
    L^\prime = L_m^\prime + L_d^\prime \approx \frac{2\sqrt{2} v_g}{\sqrt{ (\Delta \varepsilon_r / \varepsilon_r) \omega \cdot \Omega}}.
    \label{eqn:tandem_L}
\end{equation}
In this design, which we refer to as the \textit{short delay line} configuration, both the large modulation strength and the high modulation frequency can contribute to reducing the required length of the modulators.

\subsubsection{Photonic transition}
\label{sec:transition}

The spatial symmetries of modes in a waveguide are extremely important degrees of freedom in general, and can be utilized in the presence of dynamic modulation to further extend the waveguide's scattering response.
In this section we discuss isolator waveguide designs which incorporate such spatial degrees of freedom, which was first introduced in \cite{winn_interband_1999}.
The class of devices we describe here are based on an effect which is often referred to as a \textit{photonic transition} \cite{winn_interband_1999, dong_inducing_2008, yu_complete_2009}, because the modulation causes optical signal energy to transition between different optical modes, similarly to nonlinear frequency conversion in optical waveguides \cite{boyd_nonlinear_2008}.

\begin{figure*}[tb]
    \centering
    \includegraphics{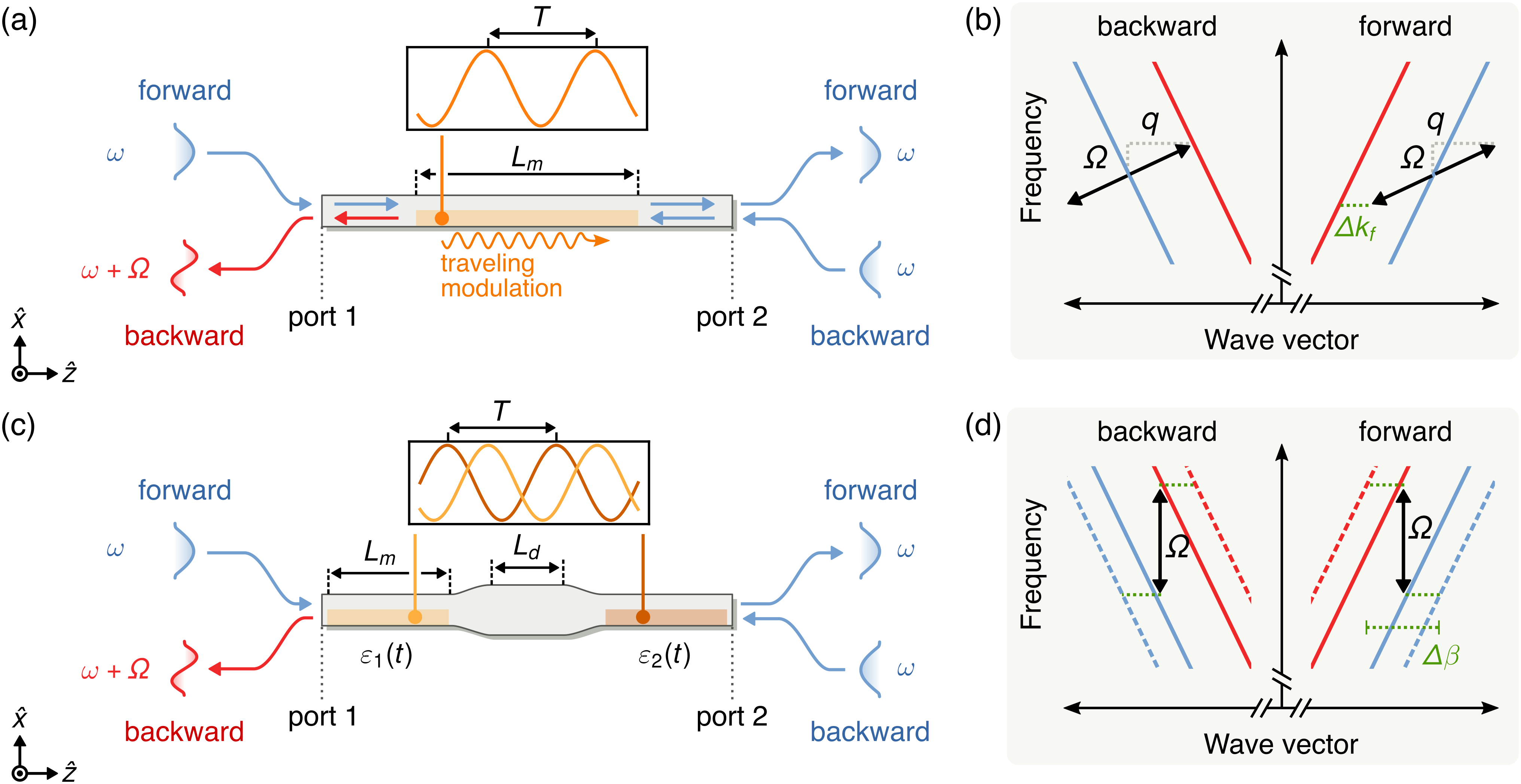}
    \caption{Optical isolator architectures based on photonic transitions acting as nonreciprocal spatial mode and frequency converters. 
    (a) Schematic and (b) waveguide mode dispersion for an isolator based on an indirect photonic transition, as proposed in \cite{yu_complete_2009}. 
    The permittivity modulation defined by (\ref{eq:eps_indirect}), with a temporal frequency of $\Omega = 2\pi/T$ and wave vector of $q$, is applied to half of the waveguide width in the transverse direction. 
    (c) Schematic and (d) waveguide mode dispersion for an isolator based on a sequence of two direct photonic transitions separated by a tapered waveguide, as proposed in \cite{fang_photonic_2012}. 
    The standing wave modulation in the direct transition has no wavevector, i.e. $q=0$, but is applied to each modulator with different phases, given by (\ref{eq:eps_direct1}) and (\ref{eq:eps_direct2}).
    The devices in (a) and (c) act as nonreciprocal spatial mode and frequency converters but spatial mode filters (not depicted) can be used to absorb or scatter the odd spatial waveguide mode from the waveguide to convert the design into an isolator.}
    \label{fig:waveguide_transition}
\end{figure*}

To review the concept of the photonic transition, we consider a model system consisting of a dielectric slab waveguide of width $w$ and relative permittivity $\varepsilon_r$, as shown in Fig \ref{fig:waveguide_transition}(a).
We focus on this system to introduce the operating principle of the device but the requirements we outline apply to realistic three-dimensional waveguides as well.
For the photonic transition, we consider the two lowest-order modes of the waveguide which have their electric fields polarized along the $y$-direction.
We sketch the dispersion band diagram for these two spatial modes, which have even and odd modal symmetry with respect to the waveguide center, in Fig. \ref{fig:waveguide_transition}(b).
In a static waveguide without modulation, these two spatial modes are uncoupled. 
However, when the permittivity of the core region is dynamically modulated, with a time-dependence of
\begin{equation}
	\frac{\varepsilon(t, x, z)}{\varepsilon_0} = \varepsilon_r + \Delta{\varepsilon_r{(x)}} \sin(\Omega t - q z)
	\label{eq:eps_indirect}
\end{equation}
the two spatial modes can couple through an indirect transition at points on their dispersion curves, $(k_1, \omega_1)$ and $(k_2, \omega_2)$, which are separated in frequency by $\Omega = \omega_2 - \omega_1$ and in wave vector by $q = k_2 - k_1$.
This process is analogous to the indirect transitions of electrons in semiconductors \cite{winn_interband_1999}.
Moreover, the requirements on the wave vector and frequency matching are mathematically equivalent to the phase matching requirements in optical sum and difference-frequency generation \cite{boyd_nonlinear_2008}.
Clearly, the form of dynamic modulation defined by (\ref{eq:eps_indirect}) breaks the generalized time-reversal symmetry of (\ref{eqn:tr_eps}), but an additional requirement for the modulation to couple the two modes is that its transverse profile, given by $\Delta{\varepsilon_r{(y)}}$, must break spatial symmetry with respect to the center of the waveguide.
There are many modulation configurations that could satisfy this requirement, but perhaps the simplest is one which modulates only half of the waveguide width in the transverse direction, as shown in Fig. \ref{fig:waveguide_transition}(a).

Conceptually, the nonreciprocal operation of the indirect transition device can be understood as follows. 
In the \textit{backward} direction, an input signal entering port 2 in the even spatial mode at frequency $\omega$ will have its energy parametrically converted to the odd spatial mode at frequency $\omega + \Omega$ as it propagates down the length of the waveguide.
By designing the length of the modulated segment, $L_m$, such that the signal is completely converted, in the \textit{backward} direction, the device acts as a spatial mode and frequency converter.
However, in the \textit{forward} direction, an input signal entering port 2 in the even spatial mode at frequency $\omega$ is unaffected by the modulation because it experiences a large phase mismatch, as indicated by $\Delta{k_f}$ on the left hand side of Fig. \ref{fig:waveguide_transition}(b).

The device described above is, essentially, a nonreciprocal spatial mode and frequency converter between four ports: the Floquet ports with $p=0$ at frequency $\omega$ at port 1 and port 2 of the waveguide (which is always associated with the even spatial mode) and the Floquet ports with $p=+1$ at port 1 and port 2 of the waveguide (which, in this configuration, is always associated with the odd spatial mode at frequency $\omega+\Omega$).
To summarize, the nonreciprocal scattering in the device takes place as follows:
\begin{itemize}
    \item The even spatial mode entering port 1 at frequency $\omega$ is transmitted to the even spatial mode with frequency $\omega$ at port 2.
    \item The even spatial mode entering port 2 at frequency $\omega$ is transmitted to the odd spatial mode with frequency $\omega + \Omega$ at port 1.
    \item The odd spatial mode entering port 1 at frequency $\omega + \Omega$ is transmitted to the odd spatial mode with frequency $\omega + \Omega$ at port 2.
    \item The odd spatial mode entering port 2 at frequency $\omega + \Omega$ is transmitted to the even spatial mode with frequency $\omega$ at port 1.
\end{itemize}
The first two bullet points outlined above correspond to the pathways illustrated in Fig. \ref{fig:waveguide_transition}(a).
To construct an isolator for an input and output signal in the even spatial mode at frequency $\omega$, a spatial mode filter can be introduced in series with the modulated waveguide to scatter or absorb the odd spatial waveguide mode.
By using such a filter, an isolator is constructed from a subset of circulator ports, as illustrated in Fig. \ref{fig:scattering}(b).

The \textit{indirect transition} described above relies entirely on the modulation wave vector, $q$, to break the generalized time reversal symmetry given by (\ref{eqn:tr_eps}).
In contrast, when $q=0$ but with a modulation frequency still satisfying $\Omega = \omega_2 - \omega_1$ for some point $(k, \omega_1)$ on the even mode band and some $(k, \omega_2)$ on the odd mode band, the coupling becomes a \textit{direct transition}.
Because this form of modulation is a standing wave, it does not break the generalized time-reversal symmetry given by (\ref{eqn:tr_eps}), and the device configuration, as shown in Fig. \ref{fig:waveguide_transition}(a), becomes reciprocal.
However, an isolator can be constructed by setting up a sequence of two direct transitions using the configuration proposed in \cite{fang_photonic_2012}, as shown in Fig. \ref{fig:waveguide_transition}(c).
This device consists of two modulated waveguide segments, that have a time-dependence of
\begin{align}
	\varepsilon_1(t, y) &= \varepsilon_r + \Delta{\varepsilon_r{(y)}} \sin(\Omega t + \phi_1) \label{eq:eps_direct1} \\
	\varepsilon_2(t, y) &= \varepsilon_r + \Delta{\varepsilon_r{(y)}} \sin(\Omega t + \phi_2). \label{eq:eps_direct2}
\end{align}
Note the different phases in the modulating waveforms. 
The two modulated waveguide segments are separated by a wider waveguide of width, $w^\prime$ and length $L_d$ which is unmodulated.
The modulated waveguides support an even and odd spatial mode with dispersion corresponding to the solid lines in Fig. \ref{fig:waveguide_transition}(d), while the wider unmodulated waveguide supports an even and odd spatial mode with dispersion corresponding to the dashed lines in Fig. \ref{fig:waveguide_transition}(d).

Conceptually, the nonreciprocal operation of the \textit{direct transition} isolator  shown in Fig. \ref{fig:waveguide_transition}(c) can be understood as follows. 
Two pathways through the center part of the device are supported: one in the even spatial mode at frequency $\omega$ and one in the odd spatial mode at frequency $\omega + \Omega$, which experience a relative phase shift of $-L_d \Delta{\beta}$ due to the dispersion shown in Fig. \ref{fig:waveguide_transition}(d).
These two pathways are coupled together by the direct transitions in the modulated waveguides which impart a $+\phi_{1,2}$ phase shift when coupling upward in frequency and a $-\phi_{1,2}$ phase shift when coupling downward in frequency.
The interference between these two pathways, which includes the direction-dependent phase shift, is what leads to a nonreciprocal response.
Specifically, in the \textit{backward} direction, the two pathways for an input entering port 2 in the even spatial mode at frequency $\omega$ interfere at port 1 with a relative phase of $\pi -L_d \Delta{\beta} + \phi_1 - \phi_2$.
In the \textit{forward} direction, the two pathways for an input at port 1 in even spatial mode at frequency $\omega$ interfere at port 2 with a relative phase of $\pi -L_d \Delta{\beta} + \phi_2 - \phi_1$.
Note the change in sign on the $\phi_1$ and $\phi_2$ terms between the forward and backward directions.
Thus, complete nonreciprocal spatial mode conversion can be achieved by configuring the phases of the modulation to satisfy $\phi_1-\phi_2 = \pi/2$ and by designing the unmodulated waveguide to have a dispersion which satisfies $L_d \Delta{\beta} = \pi/2$.
These two conditions result in constructive interference for an input signal in the even spatial mode in the \textit{forward} direction and destructive interference in the \textit{backward} direction.

Thus, the device shown in Fig. \ref{fig:waveguide_transition}(c) acts as a nonreciprocal spatial mode and frequency converter, like the device shown in Fig. \ref{fig:waveguide_transition}(a).
Similarly, it can be operated as an isolator by including a spatial mode filter which scatters or absorbs the odd spatial mode.
We emphasize that, although the standing wave modulation given by (\ref{eq:eps_direct1}) and (\ref{eq:eps_direct2}) have local time-reversal symmetry, the entire device, which includes the combination of $\varepsilon_1(t)$ and $\varepsilon_2(t)$, does break (\ref{eqn:tr_eps}).

The bandwidth and isolation of both the \textit{indirect} and \textit{direct} photonic transition isolators are determined by the dispersion of the waveguide modes \cite{yu_integrated_2010, fang_photonic_2012, williamson_broadband_2019a}.
We now discuss how the required lengths of these two devices determines their bandwidth and isolation performance.
For both the \textit{direct} and \textit{indirect} transition, the length of the modulated waveguide required for complete conversion is referred to as the coherence length. 
When the bands of the two modes are nearly parallel, i.e. $v_{g1}(\omega) \approx v_{g2}(\omega + \Omega) = v_g$, and the modulation frequency is much smaller than the optical frequency, i.e. $\Omega \ll \omega$, the coherence length is given by \cite{yu_integrated_2010}
\begin{equation}
    L_c = 2\pi\frac{v_g}{\eta \omega}, \label{eq:Lc}
\end{equation}
where $c_0$ is the speed of light and $\eta$ is the effective modulation strength defined by the overlap integral
\begin{equation}
    \eta = \frac{\int{ {{\Delta\varepsilon_r}(x)} {E_1(x)} {E_2^*(x)} dx } }{\sqrt{ \int{{\varepsilon_r(x)} |{E_1(x)}|^2 dx} \int{ {\varepsilon_r(x)} |{E_2(x)}|^2 dx} } }.
    \label{eq:transition_coupling}
\end{equation}
In (\ref{eq:transition_coupling}), ${E_1(x)}$ and ${E_2(x)}$ are the transverse spatial profiles of the even and odd waveguide modes.
As discussed above, to operate as an isolator, the modulated waveguide in the \textit{indirect} transition isolator [Fig. \ref{fig:waveguide_transition}(a)] must satisfy $L_m = L_c$ for complete conversion of energy from the even spatial mode to the odd spatial mode in the backward direction.
For the \textit{direct} transition isolator [Fig. \ref{fig:waveguide_transition}(c)], the length of each modulator must satisfy $L_m = L_c/2$ to split energy equally between the two modes.
Therefore, the total length of the device shown in Fig. \ref{fig:waveguide_transition}(a) is $L = L_m = L_c$, while the total length of the design in Fig. \ref{fig:waveguide_transition}(c) is $L = 2 L_m + L_d = L_c + L_d$, not accounting for the size of the spatial mode filters.
Here, $L_d$ is constrained by the design of the delay line waveguide and, specifically, the maximum achievable shift between the bands of the even and odd spatial mode.

The finite value of $\Delta{k_f}$ in the forward direction of the indirect transition, as shown in Fig. \ref{fig:waveguide_transition}(b), leads to the undesired conversion of signal energy into the odd spatial mode, with an efficiency proportional to $1/\left(\Delta{k_f}L_m\right)$.
Such conversion also occurs for the \textit{direct} transition, although $\Delta{k_f}$ is not explicitly indicated in Fig. \ref{fig:waveguide_transition}(d).
This signal conversion effectively acts as an insertion loss for the signal.
To limit this effect, isolators based on the photonic transition are constrained to operate with a coherence length that is large enough to make $\Delta{k_f} L_m \gg 1.$
Because $\Delta{k_f} \approx 2\Omega/v_g$, the length of the modulated waveguide is constrained to
\begin{equation}
    L_m \ge \frac{v_g}{2\Omega},
\end{equation}
which limits the modulation strength to
\begin{equation}
    \eta \le 4\pi\frac{\Omega}{\omega}.
\end{equation}

For a photonic transition isolator designed to operate with an input signal at $\omega$, the modulation satisfies the phase matching condition of ${\Delta{k}(\omega)} = q - \left[k_2(\omega+\Omega) - k_1(\omega)\right] \equiv 0$.
For an input at a nearby frequency of $\omega^\prime = \omega + {\Delta{\omega}}$, the phase matching condition may not be exactly satisfied due to group velocity dispersion.
In other words, ${\Delta{k}(\omega^\prime)} \ne 0$ for a nonzero $\Delta{\omega}$.
Such phase mismatch will result in incomplete conversion of signal energy into the odd spatial mode and results in a degraded isolation in the backward direction, with some signal energy remaining in the even spatial mode.
Therefore, the dependence of ${\Delta{k}(\omega^\prime)}L_m$ on $\Delta{\omega}$ determines the isolation bandwidth.
Additionally, dispersion in the spatial mode field profiles affects the effective modulation strength, given by (\ref{eq:transition_coupling}), and also limits the performance of the device, by changing the coherence length.

\subsubsection{Gain-loss modulation}
\label{sec:gain_loss}
Instead of applying dynamic modulation to the real part of the permittivity, as in the waveguide devices discussed in previous sections, \cite{PhysRevA.99.013824} considered a device where dynamic modulation was applied to the imaginary part of the permittivity, i.e. the gain and loss of a waveguide.
Like the photonic transition described in the previous section, the gain-loss modulated isolator also utilizes different spatial modes of the waveguide. 

The device structure considered in \cite{PhysRevA.99.013824} is similar to the one shown in Fig.~\ref{fig:waveguide_transition}(a), in which a waveguide supports an even and odd spatial mode in the absence of modulation, but instead has a conductivity with a traveling wave time-dependence of
\begin{equation}
    \sigma(t, x, z) =  \Delta\sigma(x) \sin(\Omega t - qz),
    \label{eq:modulation_profile}
\end{equation}
where $\Delta\sigma(x)$ is the conductivity modulation profile in the transverse direction of the waveguide, $\Omega$ is the modulation frequency, and $q$ is the modulation wave vector. 
The gain-loss modulation described by (\ref{eq:modulation_profile}) does not create a nonreciprocal frequency and spatial mode conversion like the indirect transition, but instead results in nonreciprocal amplification and attenuation. 

The behavior of the waveguide under gain-loss modulation extending from $z=0$ to $L_m$ can be summarized by
\begin{equation}
    \begin{bmatrix} a_e(L_m)\\a_o(L_m) \end{bmatrix} 
    = 
    \begin{bmatrix} T_{11} & T_{12} \\ T_{12}^* & T_{11}^* \end{bmatrix} 
    \begin{bmatrix} a_e(0)\\a_o(0) \end{bmatrix}
    \label{eq:gain_loss_tmatrix}
\end{equation}
where the elements of the transfer matrix are:
\begin{align}
    T_{11} &= e^{j \frac{\Delta k}{2} L_m} \left(\cosh(\eta' L_m) -  j\frac{{\Delta} k}{2\eta'} \sinh(\eta' L_m)\right) \\
    T_{12} &= -e^{j\frac{\Delta k}{2} L_m}\frac{\eta}{\eta'} \sinh(\eta'L_m),
\end{align}
with $\eta=\frac{1}{8}\int{\Delta\sigma(x)E_1(x)E_2(x)dx}$ and $\eta' = \sqrt{\eta^2 - (\Delta k/2)^2}$.
In (\ref{eq:gain_loss_tmatrix}), $a_e$ and $a_o$ are the modal amplitudes of the even waveguide spatial mode at frequency $\omega$ and the odd waveguide spatial mode at frequency $\omega+\Omega$, respectively.

Like the photonic transition isolators in the previous section, the phase mismatch, $\Delta k$ from the gain-loss modulation is different for the \textit{forward} and \textit{backward} directions, which provides the nonreciprocal response.
We first consider the case of the \textit{backward} direction where, ideally, the phase mismatch, $\Delta k$ is zero, as shown in Fig. \ref{fig:waveguide_transition}(b).
From (\ref{eq:gain_loss_tmatrix}), a signal input at frequency $\omega$ in the even spatial mode at port 2 will be amplified with a transmission of $\cosh^2(\eta L_m)$ to the even spatial mode at port 1.
Simultaneously, an output in the odd spatial mode at port 1 will be generated, with a transmission of $\sinh^2(\eta L_m)$.
In the \textit{forward} direction, due to the large phase mismatch $\Delta{k} \gg \eta$, as indicated by Fig. \ref{fig:waveguide_transition}(b), an input in the even spatial mode at frequency $\omega$ at port 1 propagates without attenuation or amplification. 
Considering that the input signal at frequency $\omega$ in the even spatial mode is amplified in the \textit{backward} direction but unchanged in the \textit{forward} direction, the waveguide under gain-loss modulation provides a nonreciprocal amplification. 

The gain-loss modulated waveguide described above can be configured either as a directional amplifier \cite{Fang2017,PhysRevX.5.021025,Li:17} or as an isolator. 
To operate as an isolator, a suitable background loss can be introduced to the modulated waveguide or an absorptive waveguide segment can be added in series with the modulated waveguide.
Such a device configuration would result in no amplitude change for a signal in the even spatial mode entering in the \textit{backward} direction.
In the \textit{forward} direction, a signal in the even spatial mode entering port 1 would experience exponential attenuation, with the absorption playing the role of the additional port shown in Fig. \ref{fig:scattering}(b).
Note that the \textit{forward} and \textit{backward} direction labels in such an isolator configuration need to be swapped in order to remain consistent with the convention used by other devices discussed in this review.

Unlike the indirect photonic transition described in the previous section, the gain-loss isolator does not have the notion of a coherence length, at which optimal isolation is achieved. 
Instead, the waveguide with gain-loss modulation achieves an isolation ratio that is proportional to its length, $L_m$.
Assuming an ideal phase matching for the \textit{backward} direction, the maximal isolation is $\text{IR}_{\text{max}} = L_m\Delta k_f/2 \approx L_m\Omega/v_g$ and is obtained when $\eta=\Delta k_f/2$.

We note that there is intrinsic spontaneous emission noise associated with optical gain \cite{caves_1982, haus2010electromagnetic}.
Additionally, at thermal equilibrium a device with loss is also subject to thermal fluctuation noise \cite{callen_1951}.
A full treatment of the transport properties of any device with gain and loss must take these sources of noise into account. 
In principle, these sources of noise can be treated by introducing additional ports to the device's scattering matrix and by associating the corresponding noise operators with each of these additional ports \cite{haus2010electromagnetic}.
However, even in the presence of such noise sources, a device with modulated gain and loss remains nonreciprocal.

\subsection{Modulated Resonators}
\label{sec:devices_resonators}

Optical resonators are devices that confine and localize optical energy, which allows for significant enhancement of the effective modulation strength and far more compact device footprints, as compared to waveguide devices. 
Unlike waveguides, resonators can also be coupled to a number of additional ports, including those associated with waveguides but also those corresponding to radiation loss and material absorption pathways.
Such forms of coupling open up additional degrees of freedom that can be used to construct dynamically modulated optical isolators and circulators.

\subsubsection{Photonic transition in a ring resonator}
\label{sec:transition_ring}

\begin{figure*}[tb]
    \centering
    \includegraphics{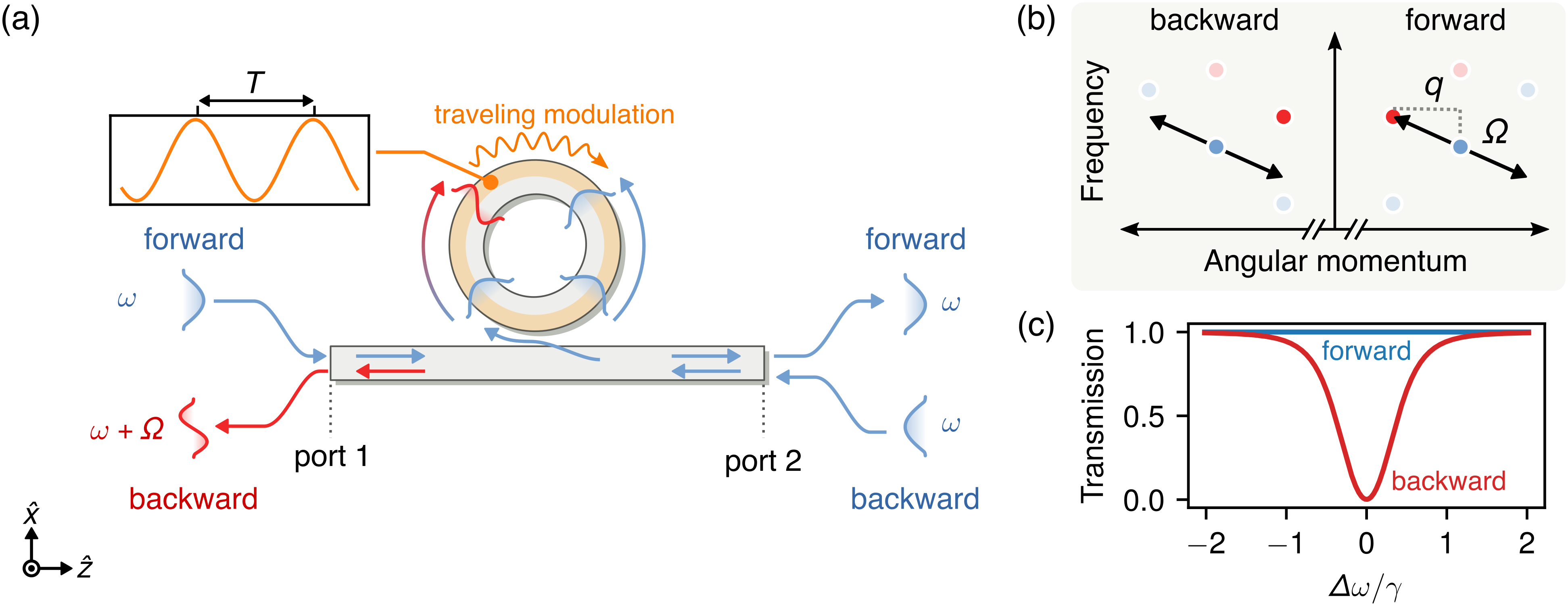}
    \caption{
    (a) Schematic of the resonant photonic transition isolator proposed in \cite{yu_complete_2009}. Both the ring resonator and the waveguide support multiple transverse spatial modes. 
    (b) Frequency-momentum diagram of the CW and CCW ring resonator modes with the modulation arrow overlaid.
    (c) Forward and backward transmission spectrum around $\omega_1$, where the detuning is $\Delta{\omega} = \omega - \omega_1$ and $\gamma = \gamma_1 + \gamma_2$ is the half-width half-max linewidth of the resonator.}
    \label{fig:resonator_transition}
\end{figure*}

In this section, we review the version of the photonic transition isolator, as shown in Fig. \ref{fig:resonator_transition}(a).
This device is a direct analogy to the nonreciprocal spatial mode and frequency converter of the photonic transition isolator shown in Fig. \ref{fig:waveguide_transition}(a). 
However, the key difference from the continuous dispersion of the waveguide, is that the ring supports resonances at discrete angular momenta and frequencies, as shown in Fig. \ref{fig:resonator_transition}(b).
In the context of an optical ring resonator, the angular momentum of a ring mode refers to the integer number of guided optical wavelengths that exist in the mode for a single round trip around the ring.

To review the operating principle of the device, we consider a ring resonator side-coupled to an access waveguide, as shown in Fig. \ref{fig:resonator_transition}(a).
The ring has a radius $R$ and is constructed from a waveguide of width $w$ and permittivity $\varepsilon_r$ and the access waveguide also has a width $w$ and permittivity $\varepsilon_r$.
For the photonic transition, we focus on two degenerate pairs of clockwise and counter-clockwise traveling-wave modes that have their electric fields polarized along the $y$-direction. 
The first resonator mode pair has a frequency $\omega_1$, an angular momentum $\pm q_1$, and an even profile with respect to the center of the ring waveguide.
The second resonator mode pair has a frequency $\omega_2$, an angular momentum $\pm q_2$, and an odd profile with respect to the center of the ring waveguide.
The ring is designed such that each resonator mode has only a \textit{single} input-output coupling channel.
Specifically, the $\omega_1$ ring resonator mode couples to the even spatial mode of the waveguide with a rate $\gamma_1$ and the $\omega_2$ ring resonator mode couples to the odd spatial mode of the waveguide with a rate $\gamma_2$.
Any loss rate of the resonator modes associated with material absorption, $\gamma_a$, or radiation, $\gamma_r$, are negligible in comparison to the waveguide-ring coupling rate, i.e. $\gamma_a, \gamma_r \ll \gamma_{1,2}$.
In a ring resonator, $\gamma_a$ can be made negligible by using a low-loss material while $\gamma_r$ can be reduced by limiting waveguide bending losses.
In practice, bending losses can be suppressed using a combination of a large ring radius and waveguide design supporting well-confined modes \cite{xiao_modeling_2007}.

To couple the two ring modes through a photonic transition, the permittivity of the ring is dynamically modulated with a time-dependence of
\begin{equation}
	\frac{\varepsilon(t, r, \phi)}{\varepsilon_0} = \varepsilon_r + \Delta{\varepsilon_r{(r)}} \sin(\Omega t - q \phi),
	\label{eq:eps_ring_indirect}
\end{equation}
where $r$ and $\phi$ are the cylindrical coordinate system defined with respect to the center of the ring. 
The angular momentum of the modulation in the ring is $q$, which is a direct analogy to the linear wave vector of the modulated waveguide in the previous section.
The profile of the modulation in the radial direction, $\Delta{\varepsilon_r{(r)}}$ must be designed to efficiently couple between the even and odd spatial modes of the ring.
For the design illustrated in Fig. \ref{fig:resonator_transition}(a), this is achieved by modulating only the outer portion of the ring.
In (\ref{eq:eps_ring_indirect}), $q$ determines how the modulation phase varies along the angular coordinate of the ring.
In a direct analogy to the phase matching process between the continuous waveguide bands in the indirect photonic transition shown in Fig. \ref{fig:waveguide_transition}(a), the photonic transition inside a ring must have a modulation frequency and angular momentum that satisfy $\Omega = \omega_2 - \omega_1$ and $q = q_2 - q_1$, respectively.
Here, the angular momentum of the dynamic modulation defined in (\ref{eq:eps_ring_indirect}) directly breaks the generalized time-reversal symmetry given by (\ref{eqn:tr_eps}).

From the two ports of the access waveguide, the nonreciprocal response is achieved as follows.
In the \textit{backward} direction, a signal entering port 2 of the waveguide in the even spatial mode at frequency $\omega = \omega_1$ couples to the counter-clockwise mode of the ring at a rate of $\gamma_1$ and is then converted to the odd counter-clockwise mode of the ring at frequency $\omega_2$ at a rate of $\eta$ as it circulates within the ring.
The signal energy at frequency $\omega_2$ then couples out of the ring to the odd spatial mode of the waveguide at a rate of $\gamma_2$.
In the \textit{forward} direction, a signal entering port 1 of the waveguide in the even spatial mode at frequency $\omega = \omega_1$ couples to the clockwise mode of the ring with a rate of $\gamma_1$.
However, because the modulation frequency and angular momentum do not match the difference between any of the clockwise mode pairs in the ring, as shown in Fig. \ref{fig:resonator_transition}(b), the signal then couples back out of the ring into the even spatial mode of the waveguide at a rate of $\gamma_1$.
Thus, the combination of the ring and access waveguide act as a nonreciprocal spatial mode and frequency converter.

Like the photonic transition devices shown in Fig. \ref{fig:waveguide_transition}, the nonreciprocal scattering takes place between four ports as follows:
\begin{itemize}
    \item The even spatial mode entering port 1 on-resonance, at frequency $\omega_1$, is transmitted to the even spatial mode with frequency $\omega_1$ at port 2.
    \item The even spatial mode entering port 2 on-resonance, at frequency $\omega_1$, is transmitted to the odd spatial mode with frequency $\omega_2 = \omega_1 + \Omega$ at port 1.
    \item The odd spatial mode entering port 1 on-resonance, at frequency $\omega_2$, is transmitted to the odd spatial mode with frequency $\omega_2$ at port 2.
    \item The odd spatial mode entering port 2 on-resonance, at frequency $\omega_2$, is transmitted to the even spatial mode with frequency $\omega_1 = \omega_2 - \Omega$ at port 1.
\end{itemize}
The first two bullet points outlined above correspond to the pathways illustrated in Fig. \ref{fig:resonator_transition}(a).
This four-port circulator response can then be converted into a two-port isolator response for a signal at $\omega = \omega_1$ by inserting a spatial mode filter in series with the device to scatter or absorb the odd spatial waveguide mode \cite{yu_complete_2009}.

Importantly, to achieve complete signal conversion in the resonant process described above, and thus complete isolation, the device must be designed such that the modulation-induced coupling rate. $\eta$, is equal to the geometric average of the ring-to-waveguide coupling rates given by
\begin{equation}
    \eta = \sqrt{\gamma_1 \gamma_2}. \label{eq:ring_trans_cc}
\end{equation}
The coupling rates between the ring and waveguide modes, i.e. $\gamma_1$ and $\gamma_2$, are determined by the proximity of the waveguide to the ring, while the conversion rate between the two modes of the ring due to the dynamic modulation, $\eta$, is proportional to an overlap integral which is defined similarly to the expression given by (\ref{eq:transition_coupling}).
We emphasize that for the photonic transition to function as an ideal isolator, it requires that each ring mode couples to only a single input-output channel.
This means that, although the waveguide supports multiple spatial modes in the frequency range of interest (with each spatial mode acting as a potential coupling channel), each ring mode must couple to one, and only one, of these spatial waveguide modes.
In practice, such one-to-one coupling could be achieved by designing the ring to phase match each of its modes to the target spatial waveguide mode \cite{luo_wdmcompatible_2014}.

The primary advantage of using a resonator, as opposed to the photonic transition described in Sec. \ref{sec:transition}, is the possibility of making the device much more compact. 
Specifically, the modulation strength places different constraints on the waveguide and resonator implementations of the photonic transition.
In the waveguide, the modulation strength determines the required length for suppressing phase-mismatched conversion channels as discussed at the end of Sec. \ref{sec:transition}.
Conversely, in the resonator, the modulation strength does not directly determine the device size but instead determines the required waveguide coupling rates, through (\ref{eq:ring_trans_cc}).
These coupling rates then directly determine the operating bandwidth of the device.
The ideal \textit{forward} and \textit{backward} transmission spectra for a signal around $\omega_1$ are shown in Fig. \ref{fig:resonator_transition}(c).

\subsubsection{Rabi splitting in a ring resonator}

In the version of the device described in the previous section, the ring was designed to couple each of its modes to a particular spatial mode of the waveguide, which allowed the device to act as a resonant nonreciprocal mode and frequency converter.
This design additionally required suppressing each mode of the ring from coupling to radiative loss channels, meaning that $\gamma_{r1} \ll \gamma_{1}$ and $\gamma_{r2} \ll \gamma_{2}$.
In \cite{shi_nonreciprocal_2018}, a different version of a dynamic ring isolator was proposed that operates with a stronger modulation and, additionally, a radiative loss channel for dissipation of light in the \textit{backward} direction.
The phrase \textit{Rabi splitting} in the name of this design refers to how the stronger modulation results in the splitting of the ring resonance frequencies exceeding the linewidth of the individual resonances.

\begin{figure*}
    \centering
    \includegraphics{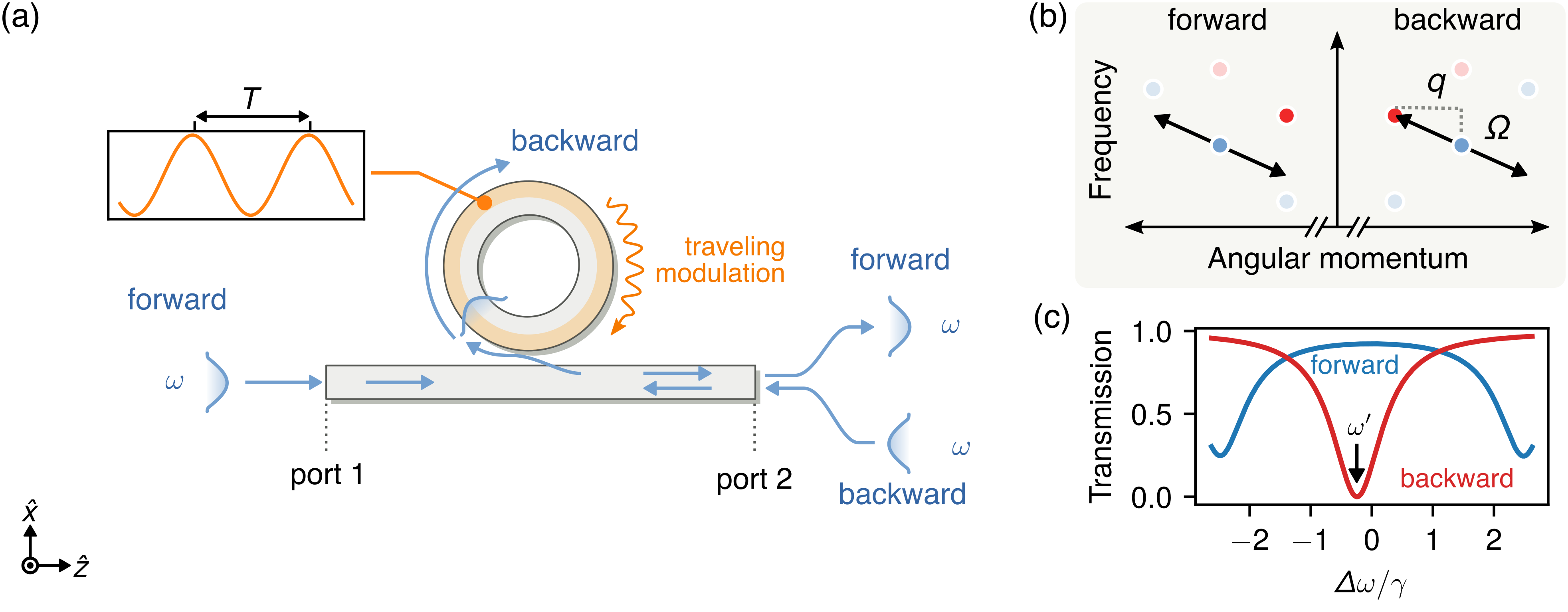}
    \caption{
    (a) Schematic of the Rabi splitting isolator proposed in \cite{shi_nonreciprocal_2018}. The ring resonator supports multiple transverse spatial modes while the waveguide supports only the fundamental transverse spatial mode. 
    (b) Frequency-momentum diagram of the CW and CCW ring resonator modes with the modulation arrow overlaid.
    (c) Forward and backward transmission spectrum around $\omega_1$, where the detuning is $\Delta{\omega} = \omega - \omega_1$ and $\gamma = \gamma_1 + \gamma_{r1}$ is the half-width half-max linewidth of the resonator.}
    \label{fig:resonator_transition_strong}
\end{figure*}

This version of the device is shown in Fig. \ref{fig:resonator_transition_strong}(a) and uses an identical ring design to the device in Fig. \ref{fig:resonator_transition}(a).
The ring is modulated with the time-dependence given by (\ref{eq:eps_ring_indirect}).
Like the photonic transition isolator, the traveling wave modulation breaks the time-reversal symmetry given by (\ref{eqn:tr_eps}).
However, unlike the design in Fig. \ref{fig:resonator_transition}(a), the strong modulation design discussed in this section uses a single-mode access waveguide that is critically coupled to the ring resonator mode at $\omega_1$, i.e. $\gamma_{r1} = \gamma_1$.
For a static ring without modulation, the critical coupling condition causes a signal entering the access waveguide from either port 1 or port 2 at a frequency $\omega = \omega_1$ to be completely dissipated into the radiation channel.
When the ring is modulated with a time-dependence given by (\ref{eq:eps_ring_indirect}) and with a large enough amplitude that the modulation-induced coupling between the two modes of the ring at $\omega_1$ and $\omega_2$ exceeds the waveguide coupling rates, i.e. $\eta > \gamma_1, \gamma_2$, the frequencies of the two ring modes shift with respect to the frequencies of the unmodulated ring modes.
The shifting observed in the resonator mode frequencies is analogous to the effect of Rabi splitting in atomic physics, also known as Autler-Townes splitting \cite{Autler1955}.

The operating principle for this device can be understood as follows \cite{shi_nonreciprocal_2018}.
In the phase-matched \textit{forward} direction, there are two resonantly coupled modes in the ring, specifically the symmetric spatial mode at $\omega_1$ and the anti-symmetric spatial mode at $\omega_2 = \omega_1 + \Omega$. 
When the modulation-induced coupling rate, $\eta$ exceeds the resonator mode linewidth, the static mode at $\omega_1$ splits into two resonances at $\omega_1 \pm \eta$. 
Therefore, in the \textit{forward} direction, a signal entering from port 1 at a frequency $\omega = \omega^\prime$ between the split resonances experiences high transmission to port 2. 
In the \textit{backward} direction, a signal entering from port 2 at a frequency $\omega = \omega^\prime$ can be completely dissipated by the loss channel of the ring.
A detailed analysis reveals the need for a careful tuning of the operating frequency, $\omega^\prime$ because the strong modulation also shifts the resonance frequencies of the modes in the \textit{backward} direction, but instead into frequencies $\omega_1 \pm \Omega$.
This can be accounted for by choosing an operating frequency for the device, $\omega^\prime$ that has complete absorption in the \textit{backward} direction, but high transmission in the \textit{forward} direction.

As noted above, a crucial difference between the Rabi splitting ring design and the photonic transition isolator design is that the \textit{forward} direction is phase-matched, rather than the \textit{backward} direction, as shown in Fig. \ref{fig:resonator_transition_strong}(b).
Additionally, whereas the photonic transition design shown in Fig. \ref{fig:resonator_transition} requires an additional spatial mode filter in the access waveguide to absorb the odd spatial waveguide mode, in the Rabi splitting design the loss channel for the signal in the \textit{backward} direction is built into the ring itself. 
Moreover, the access waveguide in this design is single-mode, meaning that the requirement in the photonic transition design required carefully engineering the coupling between the modes of the ring and the modes of the waveguide can be avoided. 
The key requirement for the Rabi splitting design is that the modulation strength satisfies the condition
\begin{equation}
    \eta > 2\gamma_1,
    \label{eq:ring_trans_strong}
\end{equation}
and that the ring resonator mode is critically coupled top the waveguide, e.g. $\gamma_1 = \gamma_{r1}$.
The ideal \textit{forward} and \textit{backward} transmission spectra for a signal around $\omega_1$ are shown in Fig. \ref{fig:resonator_transition_strong}(c).

\subsubsection{Angular momentum biasing}
\label{sec:resonator_symmetry_breaking}

In the resonant photonic transition isolator described in the previous sections, the directional coupling between the ring and waveguide plays a central role in the directional conversion process between the even and odd modes.
In this section, we review the isolator design proposed in \cite{sounas_angularmomentumbiased_2014} that breaks the degeneracy between the clockwise and counter-clockwise modes of a ring using a traveling wave modulation with a form given by (\ref{eq:eps_ring_indirect}).
Although technically all ring isolator designs that we have reviewed use the form of modulation in (\ref{eq:eps_ring_indirect}) with an angular momentum, we refer to the isolator design in this section as an \textit{Angular momentum biased} isolator because this is the name used in the original proposal \cite{sounas_angularmomentumbiased_2014}.

\begin{figure*}
    \centering
    \includegraphics{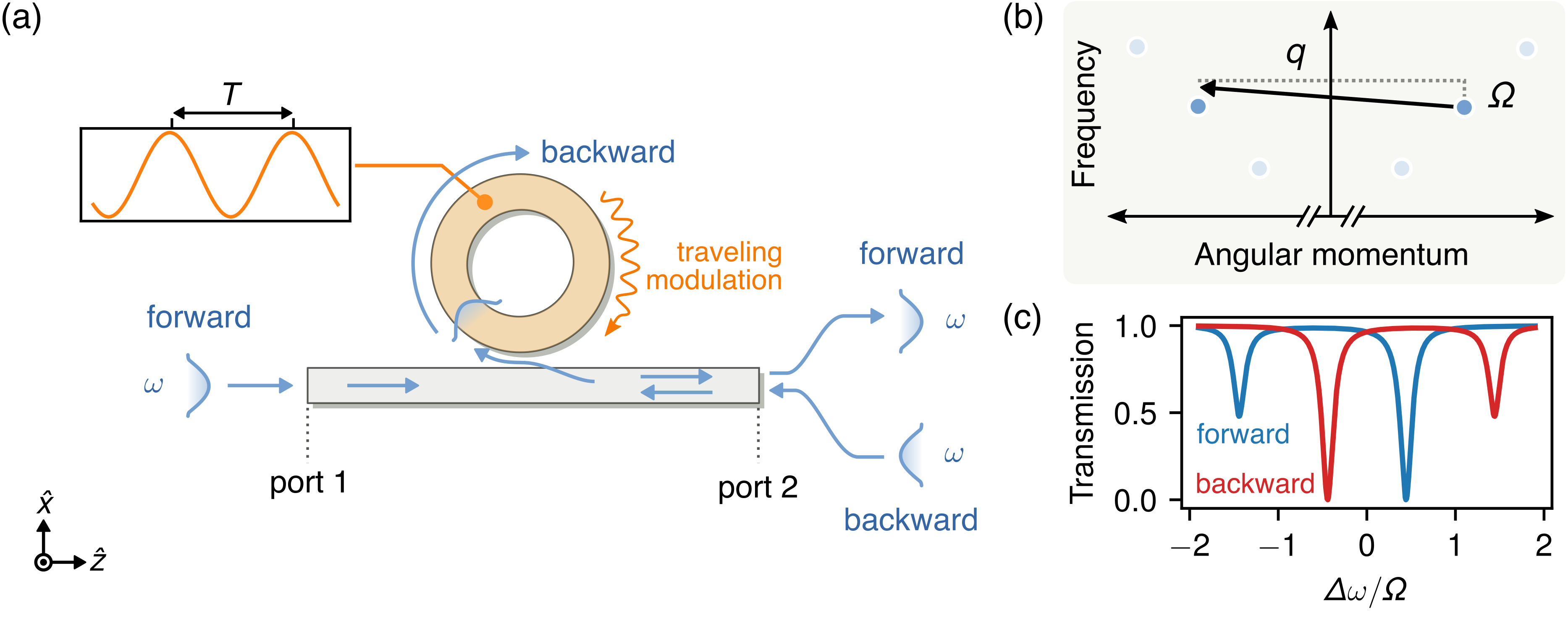}
    \caption{
    (a) Schematic of the Angular momentum biased isolator proposed in \cite{sounas_angularmomentumbiased_2014}.
    Both the waveguide and the ring resonator support only the fundamental transverse spatial mode.
    (b) Frequency-momentum diagram of the CW and CCW ring resonator modes with the modulation arrow overlaid.
    (c) Forward and backward transmission spectrum around $\omega_1$, where the detuning is $\Delta{\omega} = \omega - \omega_1$.}
    \label{fig:resonator_broken}
\end{figure*}

Unlike the resonant photonic transition isolator shown in Fig. \ref{fig:resonator_transition}, the angular momentum biased isolator shown in Fig. \ref{fig:resonator_broken}(a) involves no frequency conversion for signals in either the \textit{forward} or \textit{backward} directions.
Instead, the operating principle can be conceptually understood as being similar to magnetically-biased ring resonators.
When ring resonators with magneto-optically active materials are biased by a magnetic field, the clockwise and counter-clockwise modal degeneracy is broken because an effective optical path length difference between the clockwise and counter-clockwise modes is induced \cite{bi_onchip_2011}.
In the case of the dynamically modulated ring, the clockwise and counter-clockwise modal degeneracy is broken by the traveling wave modulation, which breaks the time-reversal symmetry condition given by (\ref{eqn:tr_eps}).

To introduce the concept for the device we consider the geometry shown in Fig. \ref{fig:resonator_broken}(a), that consists of a ring side-coupled to an access waveguide.
The ring has a radius $R$ and is constructed from a single-mode waveguide of width $w$ and permittivity $\varepsilon_r$.
The access waveguide uses a design which is identical to the waveguide used to construct the ring.
We assume that the device is invariant in the $z$-direction and focus on a pair of degenerate clockwise and counter-clockwise modes with frequency $\omega_1$ and angular momentum $\pm q_1$, as shown in Fig. \ref{fig:resonator_broken}(b).
The ring is coupled to the access waveguide and a radiation loss channel with a rate of $\gamma_1$ and $\gamma_r$, respectively, and is designed to be critically coupled, e.g. $\gamma_1 = \gamma_r$.

For a static ring without modulation, the critical coupling condition causes a signal entering the access waveguide in both the \textit{forward} or \textit{backward} directions at a frequency $\omega = \omega_1$ to be completely disspiated into the radiation channel.
In other words, there is no transmission from port 1 to port 2 (and vice-versa) on resonance.
When the ring is dynamically modulated with a time-dependence given by (\ref{eq:eps_ring_indirect}), an angular momentum satisfying $q = -2q_1$, and a uniform modulation profile across the waveguid in the radial direction of the ring, the degeneracy between the clockwise (CW) and counter-clockwise (CCW) modes is broken \cite{sounas_angularmomentumbiased_2014}.
Under such dynamic modulation, the modes have Floquet eigenfrequencies $\omega_{\text{CW}} = \omega_1 - \delta\omega/2$ and $\omega_{\text{CCW}} = \omega_1 + \delta\omega/2$, where 
\begin{equation}
    \delta\omega = \sqrt{\Omega^2 + \eta^2} - \Omega 
    \label{eq:symmetry_domega}
\end{equation}
and $\eta$ is the effective modulation coupling coefficient determined by an overlap integral between the ring modes and the modulation profile \cite{sounas_giant_2013}.
Because the Floquet eigenfrequencies are defined modulo $\Omega$, the system exhibits resonances for incident waves at frequencies $\omega_{\text{CW}} + n \Omega$ and $\omega_{\text{CCW}} + n\Omega$, where $n$ is any integer.
The resonances nearest to $\omega_{\text{CW}}$ and $\omega_{\text{CCW}}$ thus occur at frequencies $\omega_{\text{CW}} - \Omega$ and $\omega_{\text{CCW}} + \Omega$, which excite the ring resonator mode rotating in the counter-clockwise and clockwise directions, respectively.

The nonreciprocal response of the device can be understood as follows.
In the \textit{forward} direction, a signal entering port 1 at $\omega = \omega_1 - \delta\omega/2$  will be completely transmitted to port 2 because there is no resonant counter-clockwise mode in the ring at this frequency.
However, in the \textit{backward} direction, a signal entering port 2 at $\omega = \omega_1 - \delta\omega/2$ will be resonantly coupled to the radiation channel through the clockwise mode of the ring.
Therefore, this device operates as an isolator in the same spirit of Fig. \ref{fig:scattering}(b), where the third port corresponds to a radiation pathway.
The above nonreciprocal routing relies on there being no spectral overlap between any clockwise and counter-clockwise modes, which places two important constraints on the isolator design.
First, the splitting between the original clockwise and counter-clockwise modes of the ring, defined by $\delta\omega$ in (\ref{eq:symmetry_domega}), must be large enough to exceed their spectral linewidths.
Second, the nearby Floquet resonances generated by the modulation, at $\omega_{\text{CW}} - \Omega$ and $\omega_{\text{CCW}} + \Omega$, must also be spectrally separated from the split pair of clockwise and counter-clockwise modes by more than their linewidth. 
From \cite{sounas_angularmomentumbiased_2014}, the ideal modulation frequency for achieving the largest separation between neighboring resonances with opposite rotation is
\begin{equation}
    \Omega = \frac{\eta}{2\sqrt{3}} \label{eq:ring_ang_Omega}.
\end{equation}
The minimum quality factor required for a given modulation strength is then given by
\begin{equation}
    Q_{\text{min}} = 2\sqrt{3}\frac{\omega_1}{\eta} \label{eq:ring_ang_Q}.
\end{equation}
Note that in (\ref{eq:ring_ang_Q}) and (\ref{eq:ring_ang_Omega}) we have used the convention that the quality factor is $Q = \omega_1 / 2 (\gamma_1 + \gamma_r)$, where $\gamma_{1,2}$ are half-width half-max (HWHM) linewidths and $\eta$ has angular frequency units%
\footnote{Note that a different convention was used in \cite{sounas_angularmomentumbiased_2014,sounas_giant_2013} to define the strength of the modulation coupling. In \cite{sounas_angularmomentumbiased_2014,sounas_giant_2013} the coupling was represented by $\kappa$ and its relationship to the coupling rate defined in this review, $\eta$, is $\kappa = \frac{1}{2}\frac{\eta}{\omega_1}$.}.
Finally, the operating principle of the above device assumes that the modulation is weak enough to not affect the coupling rates, $\gamma_{1,2}$.
The ideal \textit{forward} and \textit{backward} transmission spectra for a signal around $\omega_1$ are shown in Fig. \ref{fig:resonator_broken}(c).

Note that the version of this device originally proposed in \cite{sounas_angularmomentumbiased_2014} actually incorporated a second coupling waveguide, rather than a radiation loss channel.
This modification to the design results in the device acting as a four port circulator, where the two physical ports of the second waveguide take the place of the radiation loss channel used for the configuration shown in Fig. \ref{fig:resonator_broken}(a).

\section{Modulation Mechanisms}
\label{sec:mechanisms}

\begin{figure}
    \centering
    \includegraphics{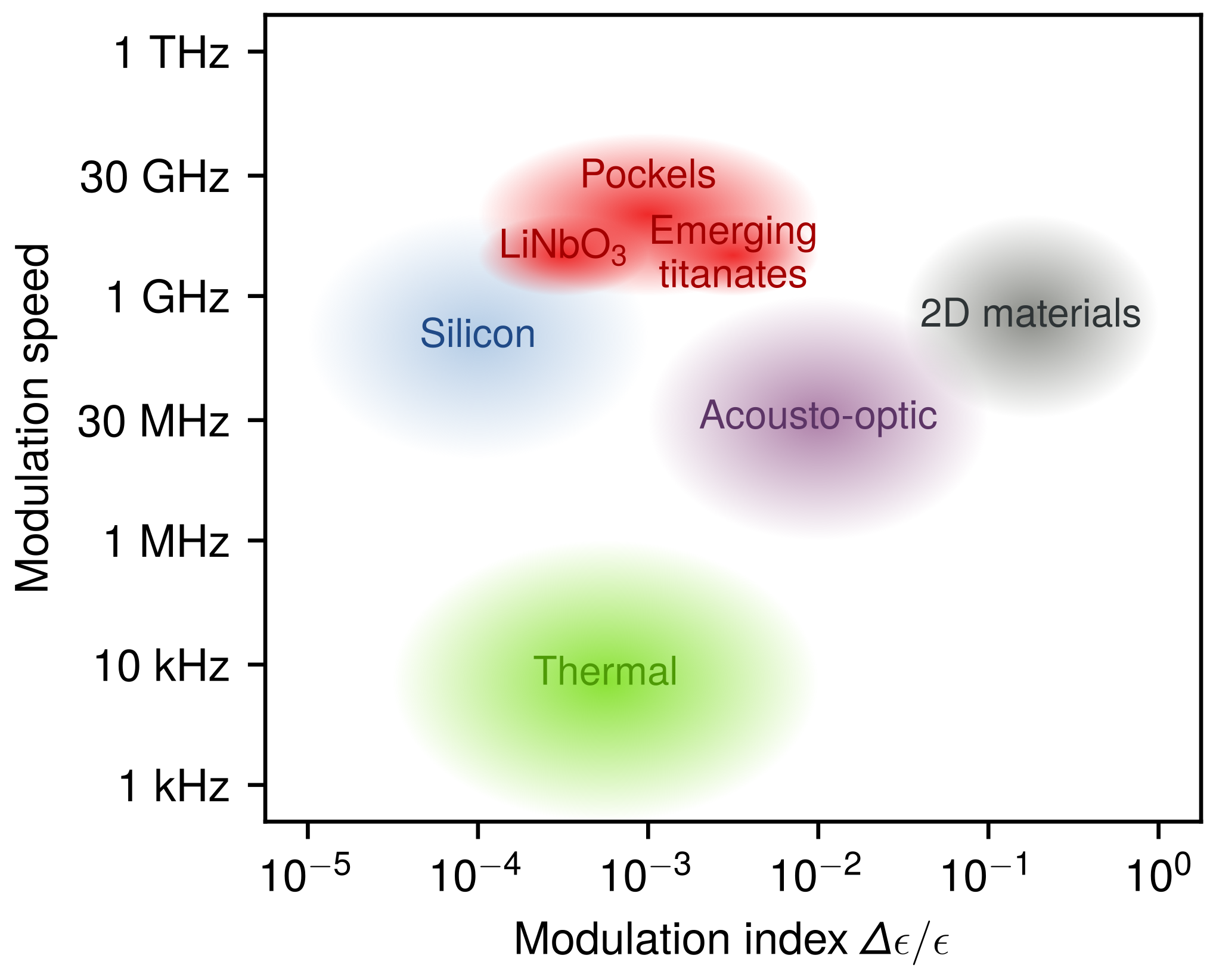}
    \caption{Comparison of modulation mechanisms in terms of their achievable speed and relative permittivity change.
    Emerging titantes include integrated barium titanate and lead zirconium titanate modulators. 
    Note that the shaded regions approximately indicate the range of best reported experimental results on chip. 
    Lower speeds and lower index modulation strengths than those indicated in the shaded regions are usually easy to achieve for each modulation mechanism, but not very useful for nonreciprocal device design.}
    \label{fig:mod_mech}
\end{figure}

In this section, we discuss relevant physical mechanisms for modulating the permittivity, focusing on material platforms that are favorable for chip-scale integration.
We discuss electro-optic (EO) modulators in traditional and emerging materials, as well as thermo-optic, acousto-optic, and all-optical modulators.
Modulation mechanisms that produce a large change in the real part of the permittivity $\Delta{\varepsilon}$ and that can operate at high speeds ($\Omega/2\pi \sim 1-100$ GHz) are ideal for implementing most of the device architectures we discuss in this review. 
We summarize the modulation speed and strength of these mechanisms in Fig. \ref{fig:mod_mech}.

Perhaps the most commonly used phase-only modulators are based on the Pockels effect, which rely on materials with a nonzero second-order nonlinearity \cite{boyd_nonlinear_2008, tsang_cavity_2010}.
Among Pockels materials, lithium niobate has been the workhorse in modulators for optical communications for decades and the recent advent of nanophotonic lithium niobate modulators in a thin-film, high-confinement geometry is extremely promising for realizing on-chip nonreciprocal devices~\cite{wang_nanophotonic_2018, wang_integrated_2018a}. 
Such nanophotonic lithium niobate modulators have achieved 40 GHz speeds with ultra low losses of 0.5 dB. 
Although III-V materials, such as gallium arsenide, have a larger second order nonlinear response than lithium niobate, high-speed on-chip EO modulators with low loss in a large-index-contrast geometry have been challenging in these materials~\cite{hiraki_heterogeneously_2017, bhandare_novel_2005a, bhasker_intensity_2018, ogiso_over_2017}. 

Lithium niobate and III-V materials are not yet compatible with CMOS technology, and hence a initial demonstrations of on-chip dynamically modulated nonreciprocal devices have focused on silicon modulators based on the plasma-dispersion effect \cite{lira_electrically_2012, tzuang_nonreciprocal_2014}. 
These demonstrations in silicon, although impressive as a proof of principle, have shown limited isolation contrast ($\sim$3 dB), and incur large insertion losses. 
The large loss arises from the use of doped regions to form the p-n and p-i-n diodes required for silicon modulators based on the plasma-dispersion effect~\cite{soref_electrooptical_1987, reed_silicon_2010, schmidt_compact_2007}.
Nevertheless, silicon modulators have pushed modulation speeds to the several tens of gigahertz range, leaving room for future improvements in the performance of silicon photonic isolators~\cite{xiao_high-speed_2013,dong_50-gb/s_2012}. 

A fundamental challenge in silicon phase modulators based on the plasma dispersion effect is the concomitant change in absorption, which causes residual amplitude modulation.
To overcome this challenge, the DC Kerr effect has been harnessed recently to produce an effective Pockels-like modulator in silicon with demonstrated speeds exceeding 5 GHz, although silicon by itself has a vanishing second order nonlinearity~\cite{timurdogan_electric_2017}.

In addition to the traditional material platforms discussed above, high-speed EO modulation has also been reported in emerging 2D materials platforms~\cite{sun_optical_2016, binder_graphene-based_2017, yu_2d_2017}, organic polymers~\cite{kieninger_ultra-high_2018} and titanates~\cite{li_strong_2019}. 
These materials are typically incorporated in the vicinity of waveguides constructed from traditional materials such as silicon or silicon nitride. 
2D materials in particular, which includes graphene and transition metal dichalcogenides, have shown a very large index change, $\Delta n \sim 1$, which is two to three orders of magnitude larger than the typically achievable index change in conventional materials~\cite{datta_lowloss_2020}. 
However, we note that the change in the effective index of the waveguide mode is much lower due to the small overlap of the 2D material monolayer with the waveguide mode. 
Demonstrated modulation speeds in 2D material based devices vary from several GHz to tens of GHz and are limited primarily by the RC time constant \cite{mousavi_kinetic_2016}.
However, in principle, such 2D material modulators, especially those based on graphene~\cite{sorianello_graphenesilicon_2018, phare_graphene_2015}, can operate at very high speeds $\sim100$ GHz. 

Alternatively, materials such as barium titanate (BTO) and lead zirconium titanate (PZT) have shown high-speed modulation speeds up to 65 GHz, and a large $\chi^{(2)}$ for BTO that is 30-50 times higher than lithium niobate~\cite{li_strong_2019, abel_large_2019}. 
Similarly, silicon-organic hybrid modulators have demonstrated strong modulation depths using low voltages, high speeds of up to tens of GHz and potentially low insertion losses by incorporating organic EO chromophores near a silicon waveguide~\cite{kieninger_ultra-high_2018}. 
A major advantage of several such emerging materials is that they can render active modulation functionalities to otherwise passive materials, i.e. those that do not support Pockels- or carrier-based modulation mechanisms.

In contrast to EO modulation schemes that can vary the index at high speeds, thermo-optic phase shifters are on the opposite end of the performance spectrum, in that they are limited to modulation speeds at or below 1 MHz \cite{harjanne_sub-s_2004, geis_submicrosecond_2004}. 
While they can produce a very high index change in nearly any dielectric material (up to $\Delta n\sim 10^{-2}$), which is larger than conventional EO materials, their low modulation speed may make them impractical for constructing dynamic isolators.

Mechanical degrees of freedom in mico- and nano-scale structures provides another route to imprint phase modulation on optical signals in waveguides via electrostriction and photoelastic effects \cite{rakich_giant_2012}. 
Such optomechanical coupling is particularly promising since large phase shifts can be produced over gigahertz bandwidths~\cite{balram2017} even in materials that do not exhibit the Pockels effect. 
Brillouin scattering has also been used to achieve nonreciprocity using the mechanical degree of freedom~\cite{huang_complete_2011, kang_reconfigurable_2011a, kittlaus_nonreciprocal_2018a, dong_brillouin-scattering-induced_2015}. 
Additionally, recent work shows that the mechanically mediated phase modulation~\cite{vanlaer_electrical_2018} can be significantly more efficient than the DC Kerr effect-based modulation~\cite{timurdogan_electric_2017} in suitably designed silicon photonic circuits.
The combination of large phase modulation efficiencies, gigahertz bandwidths and large wave vectors together make mechanically mediated modulation mechanisms very attractive for future progress in silicon photonic isolators.

Finally, all-optical modulation techniques can be used to change the refractive index at ultrafast speeds using Kerr-based cross-phase modulation~\cite{yu_all-optical_2016} or by injecting light-induced carriers into a material~\cite{matsuda_all-optical_2009, boyraz_all_2004}. All-optical modulation often requires high-power or pulsed optical beams, and is not straightforward to integrate in planar photonic circuits. However, benefits such as low loss, broad-wavelength operation at very high speeds~\cite{hochberg_terahertz_2006} and the possibility of signal processing entirely in the photonic domain have sustained interest in all-optical modulators. The aforementioned 2D material-based EO modulators can also produce superior performance when used for all-optical modulation~\cite{sun_optical_2016, lu_all-optical_2017}, although demonstrations have been primarily focused on off-chip geometries till date~\cite{yu_all-optical_2016}. To harness all-optical modulation for integrated nonreciprocal devices, it will be important to incorporate such techniques into planar photonic circuits.

While the mechanisms reviewed above have focused on varying the real part of the permittivity, the gain-loss isolator architecture \cite{PhysRevA.99.013824} that we have reviewed in Sec. \ref{sec:gain_loss} relies on dynamic modulation of the imaginary part of the permittivity. 
In a III-V semiconductor laser or amplifier structure, the mechanism of gain-loss modulation is built-in, i.e. by tuning the pumping level to the laser waveguide either electronically or optically.
The gain coefficient in contemporary semiconductor lasers typically reaches well over  $5\times 10^3\,{\textrm {cm}}^{-1}$ \cite{Ma2013,Chow1997}, corresponding to a large gain-loss modulation strength of $\text{Im}(\Delta\varepsilon) / \varepsilon \approx \Delta\sigma/\omega\varepsilon \geq 0.1 $, with achievable modulation speeds above 50 GHz \cite{Nakahara2015,Simoyama:12}.
Since directly modulated lasers switch between normal operation and off, modulation strength in the gain and loss reasonably reaches $\sim$ 10$^3$~cm$^{-1}$.
In these active devices, gain-loss modulation can in principle be directly integrated as a section of the waveguide, to perform functionalities such as directional amplification or isolation. 

Because dynamic isolators are active, they necessarily consume power in order to apply the modulation that breaks reciprocity. 
This aspect of their performance is in contrast to isolators that are constructed from magneto-optical materials that are biased using permanent magnets and, therefore, consume no power to break reciprocity.
However, a number of optical modulators with very high efficiency have been realized using the mechanisms reviewed above.
In particular, recent demonstrations of nanophotonic lithium niobate modulators have achieved switching energies of 37 aJ / bit \cite{wang_integrated_2018a}.
Assuming a modulation rate of 10 GHz would result in a power of 0.36 $\mu$W.
The 2D material modulators are also particularly impressive for their low energy consumption, with recent demonstrations consuming only 0.64 nW \cite{datta_lowloss_2020}.

\section{Comparison}
\label{sec:comparison}

In the previous sections we have reviewed a number of optical isolator architectures based on dynamically modulated waveguides and resonators as well as different modulation mechanisms that are available for realizing such devices.
In this section, we will compare the characteristics and performance of these devices based on the figures of merit outlined in Sec. \ref{sec:foms}.
We recall that an ideal isolator should provide complete isolation between the forward and backward directions, complete signal transmission (with no insertion loss) in the forward direction, and a broad bandwidth.
Additionally, large-scale integrated photonic circuits favor components with compact footprints to facilitate dense integration.
Here we will examine exactly how close each isolator design comes to achieving ideal performance, highlighting the tradeoffs made in each approach.

Our discussion in this section first focuses on comparing the footprint and bandwidth of each isolator design while assuming that the conditions for ideal isolation and insertion loss are satisfied.
We then discuss and compare specific factors in each isolator design that limit the isolation and insertion loss performance, while commenting on other relevant figures of merit outlined in Sec. \ref{sec:foms}.

\subsection{Footprint}

We begin our comparison in this section by discussing the required footprint of the different dynamic isolator architectures.
Generally, there are very different size requirements for resonator and waveguide isolators, resulting from the unique dynamics at play in each type of device.
Waveguides are traveling-wave devices where optical signals spend only a brief instant in a given region of the device.
Resonators, on the other hand, confine and trap light, allowing signals to spend, potentially, a very large number of optical cycles in a relatively small device region.
For example, microring resonators \cite{spencer2012integrated} and photonic crystal defect cavities \cite{sekoguchi2014photonic} can have quality factors greater than $10^6$.
Thus, resonant devices have the ability to significantly enhance the \textit{effective} modulation strength that they provide relative to the modulation strength naturally available in a given material.
In doing so, resonators can operate with a far smaller device footprint.
Such enhancement is important for dynamic isolators because, as shown in Fig. \ref{fig:mod_mech}, the achievable modulation strength of most materials is far below unity at optical frequencies, i.e. $\Delta\varepsilon/\varepsilon \ll 1$.

\begin{table*}[tb]
    \caption{Minimum design length of dynamic waveguide isolators}
    \setlength{\tabcolsep}{10pt}
    \renewcommand{\arraystretch}{1.3}
    \centering
    \begin{tabular}{|l c c l|}
        \hline
        \textbf{Device} & \textbf{Reference} & \textbf{Length dependence on system parameters} & \textbf{Achievable length}$^\dagger$  \\ 
        \hline
        Direct and indirect transition & \cite{yu_complete_2009, fang_photonic_2012} & \( \frac{\pi v_g}{\omega \cdot{{\Delta\varepsilon}/\varepsilon}}\) & 1 cm $-$ 3 mm\\
        Tandem (long delay line) & \cite{doerr_optical_2011}    & \(\frac{\pi v_g}{2\Omega }\) & 3 mm\\
        Tandem (short delay line) & \cite{lin_compact_2019}    & \(\frac{2\sqrt{2}v_g}{\sqrt{\omega \cdot \Omega \cdot {{\Delta\varepsilon}/\varepsilon}}}\) & 3 mm $-$ 300 $\mu$m \\
        \hline
    \end{tabular}
    \begin{flushleft}
        \footnotesize{%
        $^\dagger$ The achievable lengths assume a modulation frequency of $\Omega/2\pi \sim$ 10 GHz and a range of modulation strengths.
        }
    \end{flushleft}
    \label{tab:scaling_waveguide}
\end{table*}

\begin{figure}[tb]
    \centering
    \includegraphics{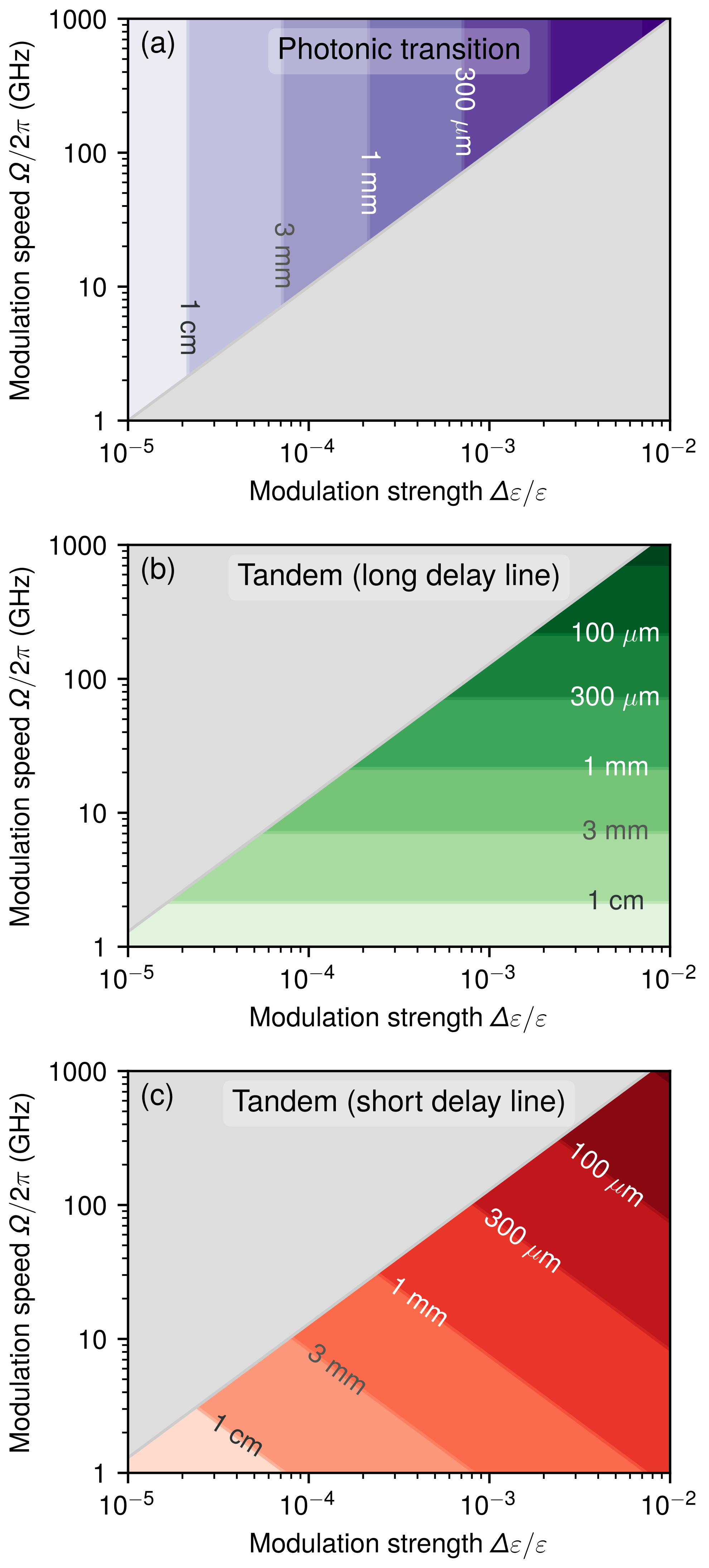}
    \caption{Minimum device length as a function of modulation strength, $\Delta{\varepsilon}/\varepsilon$ and modulation frequency, $\Omega/2\pi$ for (a) the photonic transition isolators \cite{yu_complete_2009, fang_photonic_2012} [Fig. \ref{fig:waveguide_transition}(a,c)] and the tandem phase modulator isolator [Fig. \ref{fig:waveguide_tandem}] in (b) the long delay line \cite{doerr_optical_2011} and (c) short delay line \cite{lin_compact_2019} configurations.
    The optical frequency and group velocity are assumed to be $\omega = 2\pi\cdot200$ THz and $v_g = c_0/3.5$.}
    \label{fig:scaling_waveguide}
\end{figure}

The lower bound on the size of a resonant isolator will be determined by the minimum required size of the resonator to achieve a particular quality factor, which depends on the confinement mechanism being used.
For example, ring resonators use total internal reflection in waveguides to confine light.
This means that the minimum size of a ring resonator will be constrained by waveguide bending losses, which are inversely proportional to the ring's radius.
The enhancement of modulation and size reduction in resonant devices comes, of course, with a major tradeoff for signal bandwidth, which we discuss in the next section.

For the remainder of this section, we focus on the footprint of the dynamic isolators based on waveguides, which have tight design constraints on their sizes.
Moreover, the waveguide isolator designs we have reviewed are able to operate only in specific ranges of modulation frequency and modulation strength.
Specifically, the tandem modulator (Fig. \ref{fig:waveguide_tandem}) operates in the so-called \textit{strong} modulation regime, where $\Omega \le \omega \frac{\Delta{\varepsilon}}{\varepsilon}$, while the photonic transition (Fig. \ref{fig:waveguide_transition}) operates in the \textit{weak} modulation regime, where $\Omega \gg \omega \frac{\Delta{\varepsilon}}{\varepsilon}$.
In these two inequalities, the strength of the modulation is characterized in terms of the effective change of the optical mode frequency, which is then compared to the frequency of modulation. 

The operating regimes of the waveguide-based photonic transition and the tandem modulator are complementary \cite{lin_compact_2019}, resulting from the unique approach that each architecture takes to achieve isolation.
The tandem modulator relies on strong modulation, in conjunction with a time-delay of the modulating waveform, in order to convert \textit{all} input signal energy to higher and lower frequencies in the \textit{backward} direction.
In contrast, the waveguide photonic transition relies on a phase-matched conversion process to achieve isolation in the \textit{backward} direction and, crucially, the suppression of all other phase-mismatched conversion channels.
The requirement of suppressing conversion to phase mismatched frequencies and spatial modes essentially translates into a requirement for weak modulation that facilitates a gradual conversion of energy between the even and odd modes as the signal propagates down the length of the modulator.

The minimum design lengths of the isolators are summarized in Table \ref{tab:scaling_waveguide} and are plotted in Fig. \ref{fig:scaling_waveguide}(a)-(c) as a function of the modulation frequency and modulation strength, similarly to the analysis performed in \cite{lin_compact_2019}.
The grey region of each plot in Fig. \ref{fig:scaling_waveguide} corresponds to the regime where each particular isolator design is unable to operate, e.g. the strong modulation regime for the photonic transition and the weak modulation regime for the tandem modulator isolator.
Figure \ref{fig:scaling_waveguide}(a)-(b) indicate that, with a modulation frequency of approximately 10 GHz, both the photonic transition \cite{yu_complete_2009, fang_photonic_2012} and the long delay line tandem isolator \cite{doerr_optical_2011} are restricted to a minimum device size on the order of 3 mm.
We note that this length and modulation frequency is comparable to experimentally demonstrated electro-optic phase modulators fabricated in thin film LiNbO$_3$~\cite{wang_nanophotonic_2018}.
For comparison, we note that recent experimentally demonstrated interferometric (non-resonant) magneto-optical isolators have footprints ranging from 1 - 3 mm$^2$~\cite{zhang_monolithic_2019}.

In order to achieve a more compact device, both the tandem modulator and photonic transition isolator require a simultaneously higher modulation frequency and modulation strength, corresponding to the upper right corners of Fig. \ref{fig:scaling_waveguide}(a,b).
Based on the survey of modulation mechanisms shown in Fig. \ref{fig:mod_mech}, simultaneously scaling up \textit{both} of these parameters is challenging with existing modulation mechanisms.
However, the tandem isolator can achieve a more compact device size using the short delay line configuration that was proposed in \cite{lin_compact_2019}.
This design of the tandem isolator allows the footprint to be reduced by trading off for \textit{only} an increase in the modulation strength, due to the design's $\sim 1/\sqrt{\Delta{\varepsilon}/\varepsilon}$ dependence in the modulation length (Table \ref{tab:scaling_waveguide}).
The minimum design length for this design is plotted in Fig. \ref{fig:scaling_waveguide}(c) and confirms that, for a modulation frequency of 10 GHz, a modulation strength on the order of $10^{-2}$ allows the design to accommodate a device length of approximately 300 $\mu$m, which is approximately $3\times$ smaller than the waveguide photonic transition isolator and the long delay line tandem isolator.
From Fig. \ref{fig:mod_mech}, we note that such modulation requirements could be met by a BTO electro-optic modulator \cite{li_strong_2019} or, potentially, a modulator that incorporates 2D materials.
Overall, for highly compact on-chip waveguide isolators the short delay line configuration of the tandem modulator design \cite{lin_compact_2019} is likely to be the most favorable.

Unlike the tandem modulator and photonic transition isolators, the length of the modulated gain-loss isolator depends on the target isolation ratio (IR) because the device exponentially attenuates an input signal propagating in the \textit{backward} direction.
In the gain-loss isolator, the required modulator length is $L_m = \text{IR} \cdot v_g / \Omega$.
Thus, to achieve an isolation ratio per length of 15 dB/mm (equivalent to an isolation ratio of 30 dB in a 2 mm long device) at an operating wavelength of 1.55 $\mu$m, would require a modulation frequency of 50 GHz \cite{Nakahara2015, Simoyama:12} and a modulation strength of $\text{Im}(\Delta\varepsilon)/\varepsilon\approx 1.0\times 10^{-3}$ \cite{Ma2013, Chow1997}, assuming ideal phase-matching in the \textit{backward} direction and $\Delta k_b \approx \frac{2v_g}{\Omega}$ = 72 cm$^{-1}$ in the \textit{forward} direction.

\subsection{Bandwidth}

We continue our comparison of the different dynamic isolator architectures by discussing their operating bandwidths.
A general constraint for every isolator, with the exception of the photonic transition in a waveguide (Fig. \ref{fig:waveguide_transition}), is that the isolation bandwidth is limited by the modulation frequency, $\Omega$.
An even tighter constraint for resonant dynamic isolators is that their isolation bandwidth is limited by the resonator linewidth, which is constrained to be smaller than the modulation frequency.
In this section, we first discuss the bandwidth limitation of waveguide isolators and then compare the upper bound  on the linewidth of each resonant isolator design.

The photonic transition is able to support a larger bandwidth than $\Omega$ in both the \textit{forward} and \textit{backward} directions because it uses dynamic modulation to couple different spatial modes, rather than just different frequencies.
As described in Sec. \ref{sec:transition}, the only limiting factor for the isolation bandwidth of the photonic transition is the dispersion of the optical modes. 
In particular, the necessary conditions for achieving isolation is that (1) the phase matching condition in the \textit{backward} direction, given by ${\Delta{k}(\omega)} = q - \left[k_2(\omega+\Omega) - k_1(\omega)\right] \equiv 0$, be satisfied and that (2) any other phase mismatched coupling processes be suppressed by the long length of the modulator.
Therefore, if a waveguide can be designed to support modes with parallel dispersion over some frequency range, $\Delta{\omega}$ then the waveguide can, in principle, achieve complete isolation over that same bandwidth.
In practice, achieving such parallel dispersion over a broad bandwidth may require careful engineering of the waveguide geometry \cite{williamson_broadband_2019a}. 
A similar analysis also applies to the gain-loss modulated isolator design. 
However, the material gain bandwidth in typical semiconductor diode lasers is less than 5\% of the center wavelength, which could be a tighter constraint on the bandwidth of the device than the dispersion. 
The experimental demonstration of the indirect photonic transition isolator in \cite{lira_electrically_2012} achieved nonreciprocity over a bandwidth of 200 GHz, while the analysis from \cite{poulton_design_2012} predicted that waveguide dispersion engineering could enlarge the bandwidth to 3 THz.

In contrast, the tandem modulator architecture, as shown in Fig. \ref{fig:waveguide_tandem}, always has an isolation bandwidth that is less than the modulation frequency.
We emphasize that the \textit{forward} transmission bandwidth of the tandem isolator is also limited by $\Omega$ due to the creation of intermediate modulation tones between the two modulators.
For signal bandwidths larger than $\Omega$, the generation of these intermediate tones would distort signals transmitting in the \textit{forward} direction.
Despite this limitation, we note from Fig. \ref{fig:mod_mech} that state-of-the-art on-chip Pockels modulators, based on e.g. LiNbO$_3$ or BTO, could still provide an isolator with a bandwidth greater than 10 GHz.

A modification of the tandem modulator design allows its bandwidth to be extended, up to multiples of $\Omega$, by adopting a design with parallel modulator arms \cite{doerr_silicon_2014}.
The purpose of the additional modulator arms is to cancel more of the sideband terms depicted in the top right panel of Fig. \ref{fig:waveguide_tandem}.
However, the reliance of this approach on interferometric cancellation of the modulation tones (e.g. at $\omega \pm \Omega$, $\omega \pm 2\Omega$, \ldots) in the optical domain means that non-idealities in the device may lead to incomplete cancellation of the sidebands, and therefore distortion of signals transmitting in the \textit{forward} direction.
\begin{table}[tb]
    \caption{Modulation strength - quality factor product for ring isolator designs}
    \setlength{\tabcolsep}{10pt}
    \renewcommand{\arraystretch}{1.3}
    \centering
    \begin{tabular}{|l c c|}
        \hline
        \textbf{Device} & \textbf{Reference} & $Q \cdot {\Delta\varepsilon} / \varepsilon$ $^\dagger$ \\ 
        \hline
        Photonic transition$^*$ & \cite{yu_complete_2009} & $2$ \\
        Angular momentum bias & \cite{sounas_angularmomentumbiased_2014} & $2\sqrt{3}$ \\ 
        Rabi splitting ring & \cite{shi_nonreciprocal_2018} & $4$ \\
        \hline
    \end{tabular}
    \begin{flushleft}
        \footnotesize{%
        $^\dagger$ The values reported here are for ideally coupled modes. 
        This corresponds to an ideal traveling wave modulation covering the entire waveguide. 
        For the photonic transition, this also means that each half of the waveguide is modulated with opposite polarity in order to maximally couple the even and odd modes. 
        By modulating only half of the waveguide, as shown in Fig. \ref{fig:resonator_transition}, the figure of merit $Q \cdot {\Delta\varepsilon} / \varepsilon$ is increased by a factor of approximately 2.
        } \\
        \footnotesize{%
        $^*$ For the photonic transition isolator, the geometric average of the mode quality factors, $Q_{\text{avg}} = \sqrt{Q_1Q_2}$ is used to compute $Q \cdot {\Delta\varepsilon} / \varepsilon$
        }
    \end{flushleft}
    \label{tab:scaling_bw_resonator}
\end{table}

Having discussed the bandwidth limitations of dynamic waveguide isolators, we now focus our attention on the resonant isolator designs which, as mentioned above, have tighter bandwidth constraints than the waveguide isolators.
In particular, the photonic transition and the Rabi splitting ring isolators (Fig. \ref{fig:resonator_transition}) must be designed to have linewidths that satisfy the constraint given by (\ref{eq:ring_trans_strong}) and by (\ref{eq:ring_trans_cc}), respectively.
Similarly, the angular momentum-biased ring isolator design (Fig. \ref{fig:resonator_broken}) has a linewidth constrained by (\ref{eq:ring_ang_Q}).
Despite their narrower bandwidths, these resonant devices can be far more compact than their waveguide counterparts, making them highly attractive for integrated photonic platforms.

It turns out that the bandwidth of all three ring isolator designs can be characterized by a single figure of merit: the product of the quality factor and the modulation strength, $Q\Delta{\varepsilon}/\varepsilon$.
This figure of merit was initially proposed in \cite{sounas_angularmomentumbiased_2014} and defines the quality factor, or equivalently the maximal bandwidth, that a design can achieve for a given modulation strength.
Therefore, a small value of $Q\Delta{\varepsilon}/\varepsilon$ is favorable.
The \textit{best case} value of $Q\Delta{\varepsilon}/\varepsilon$ for each isolator design, calculated from a coupled mode theory analysis \cite{haus_fields_1984, suh_temporal_2004}, is provided in Table \ref{tab:scaling_bw_resonator}.
Here, the \textit{best case} refers to a modulation profile that maximally couples the two modes, e.g. the mode at $\omega_1$ and the mode at $\omega_2$ for the photonic transition and Rabi splitting isolator designs and the two degenerate counter-rotating modes at $\omega_1$ for the angular momentum biased design.
In all designs, the best-case coupling implies an ideal sinusoidal traveling wave modulation in the ring that perfectly matches the angular momentum difference between the modes.
In the photonic transition and Rabi splitting isolator designs, maximum coupling also requires that each half of the ring waveguide be modulated with opposite polarity to maximize coupling between the even and odd modes.
Therefore, the each value in Table \ref{tab:scaling_bw_resonator} should be considered as a theoretical optimum.

\begin{figure}[tb]
    \centering
    \includegraphics{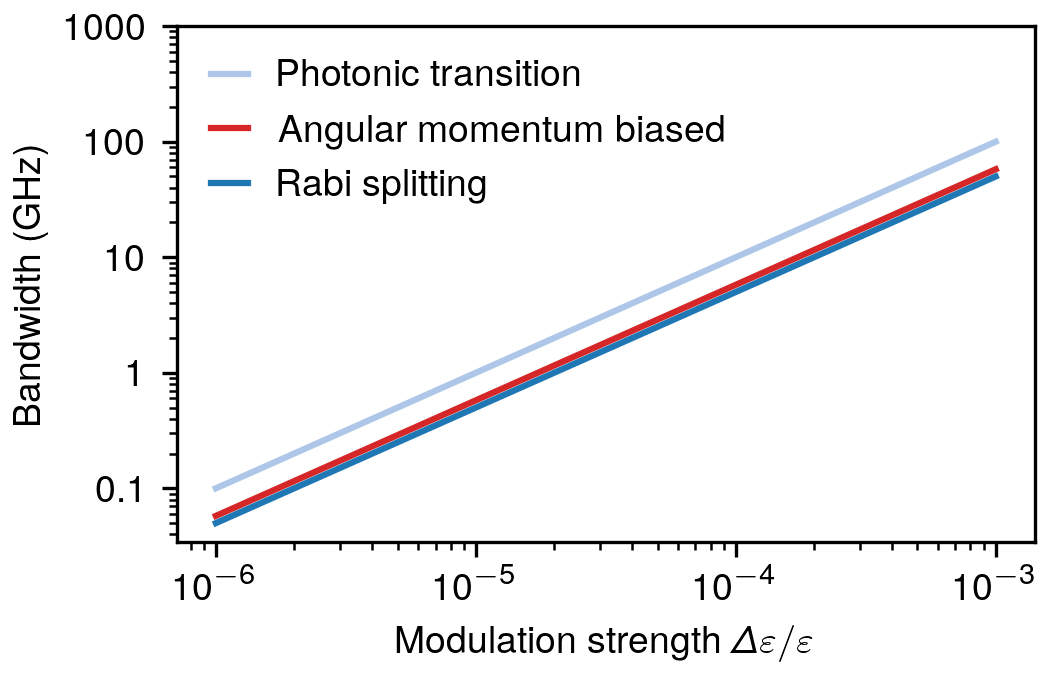}
    \caption{Operating bandwidth for photonic transition, Rabi splitting, and angular momentum biased ring resonator isolators for a given modulation strength, $\Delta{\varepsilon}/\varepsilon$. 
    This assumes an optical signal frequency of $\omega/2\pi = 200$ THz.}
    \label{fig:scaling_bw_resonator}
\end{figure}

The upper bound on the resonator linewidth for each design, calculated through $Q\Delta{\varepsilon}/\varepsilon$ and the values Table \ref{tab:scaling_bw_resonator}, is plotted as a function of the modulation strength in Fig. \ref{fig:scaling_bw_resonator}.
Here, we assume an operating frequency of $\omega/2\pi =$ 200 THz for the optical wave.
The bandwidths of the designs all scale with the same dependence on ${\Delta\varepsilon}/\varepsilon$, the Rabi splitting ring design and the angular momentum biasing design have a comparable upper bound on their bandwidth, which is approximately a factor of approximately $2\times$ smaller than the upper bound on the bandwidth of the photonic transition isolator.
We note also that the modulation frequency used for each design will need to be at least as large as the bandwidth shown in Fig. \ref{fig:scaling_bw_resonator}.

We now discuss several practical considerations that will further limit the \textit{effective} modulation strength, increase $Q\Delta{\varepsilon}/\varepsilon$, and limit the bandwidth of the ring resonator isolators discussed above.
First, in the photonic transition design, the optimum modulation profile has opposite polarity in each half of the ring waveguide. Such a modulation profile could be challenging to implement, especially if integrated metallic electrodes are involved. 
Thus, a much simpler modulation scheme for the photonic transition and the Rabi splitting isolators, as depicted in Fig. \ref{fig:resonator_transition}(a) and Fig. \ref{fig:resonator_transition}(c), is to modulate only half of the cross-section of the ring waveguide.
Note that, based on the design requirements of the angular momentum biased ring, it will not have a reduction in the effective modulation strength because it couples two modes of the same symmetry and, therefore does not need to modulate only half of the ring waveguide.
Although much more practical to fabricate, an effect of modulating only half of the ring waveguide is to immediately decrease the modulation efficiency and double the $Q\Delta{\varepsilon}/\varepsilon$ figure of merit.
Therefore, under such a modulation scheme the photonic transition design and the angular momentum biasing design have a comparable upper bound on their bandwidth, while the Rabi splitting design is approximately $2\times$ worse.

The second issue, which applies to all three ring isolator designs, is that achieving an \textit{ideal} sinusoidal traveling wave modulation, as defined in (\ref{eq:eps_ring_indirect}), is difficult using electro-optic modulation because the angular momentum of the modulation must be very large.
Here we will briefly discuss the effect that ``discretizing'' the traveling wave modulation has on the $Q\Delta{\varepsilon}/\varepsilon$ figure of merit.
Discretization of the momentum means that $N$ regions of standing-wave modulation with discrete relative phases are used to approximate the traveling wave modulation.
This approach has been theoretically proposed for all three ring isolator designs \cite{yu_complete_2009, sounas_angularmomentumbiased_2014, shi_nonreciprocal_2018} and has also been used for the experimental implementation of a waveguide-based photonic transition isolator in \cite{lira_electrically_2012}.
From the Fourier analysis presented in \cite{sounas_angularmomentumbiased_2014}, the effect of modulation discretization is to distribute the modulation over many spatial frequency components.
However, the modulation defined by (\ref{eq:eps_ring_indirect}) indicates that only a single spatial frequency component of the modulation, with wave vector of $q$, can contribute to the nonreciprocal coupling between the ring modes.
Therefore, modulation discretization reduces the effective modulation strength of the device, with an efficiency given by 
\begin{equation}
    \frac{\Delta{\varepsilon}_{\text{eff}}}{\Delta{\varepsilon}} = \text{sinc}\hspace{-2pt}\left(\frac{\Delta{M}}{ N}\right),
    \label{eq:discr_mod}
\end{equation}
where $\Delta{M}$ is the difference between the angular momentum of the coupled optical modes \cite{sounas_angularmomentumbiased_2014}.

In comparing the design requirements for each ring isolator, we note a significant difference between the modulation momentum required for each design.
The angular momentum biased ring always uses modulation to couple two optical modes with opposite rotation, which means that $\Delta{M} \equiv 2 M_1$, for an optical ring resonator mode with angular momentum $M_1$.
In other words, the angular momentum biased ring ideally requires a modulation with spatial variation on the order of half the optical wavelength.
For well-confined optical modes with high quality factors, the order of magnitude of $M_1$ is likely to be at least 10.
As demonstrated in \cite{sounas_angularmomentumbiased_2014}, the modulation can be discretized down to a configuration that uses a minimum of $N=3$ modulated regions. 
However, such a design comes with approximately an order of magnitude reduction in the effective modulation strength.
In contrast, the photonic transition isolator design can couple two optical modes with a far smaller $\Delta{M}$, meaning that the reduction in the effective modulation strength can be far smaller than in the angular momentum biased ring for an equivalent number of modulated regions, $N$.
In principle, both the photonic transition and Rabi splitting isolators could be designed to couple between two modes with $\Delta{M} = 1$.
The ability of the photonic transition and Rabi splitting isolators to operate with a far smaller $\Delta{M}$ will more than compensate for the penalty of $1/2$ that comes from modulating only half of the ring waveguide cross-section, as discussed above.
Therefore, the photonic transition designs (Fig. \ref{fig:resonator_transition}) may be the most favorable for achieving the maximum bandwidth in a compact resonant device for a given $\Delta{\varepsilon}/\varepsilon$.

\begin{table*}[tb]
    \caption{Isolator bandwidth summary}
    \setlength{\tabcolsep}{10pt}
    \renewcommand{\arraystretch}{1.3}
    \centering
    \begin{tabular}{|l c c c c c|}
        \hline
        \multirow{2}{*}{\textbf{Device}} & \multirow{2}{*}{\textbf{References}}  & \multicolumn{2}{c}{\textbf{Isolation bandwidth}} & \multicolumn{2}{c|}{\textbf{Forward bandwidth}} \\ 
         & & \textbf{Parameter} & \textbf{Achievable} & \textbf{Parameter} & \textbf{Achievable} \\
        \hline
        \textbf{Resonators} & & & & & \\
        \hline
        Photonic transition & \cite{yu_complete_2009} & $\le\gamma$ & 3 GHz & $>\Omega$ & 10 GHz \\
        Rabi splitting & \cite{shi_nonreciprocal_2018} & $\le\gamma$ & 1.5 GHz & $\le2\eta$ & 6 GHz \\
        Angular momentum bias & \cite{sounas_angularmomentumbiased_2014} & $\le\gamma$ & 1.7 GHz & $\le\gamma$ & 1.7 GHz \\
        \hline
        \textbf{Waveguides} & & & & &\\
        \hline
        Photonic transition &\cite{yu_complete_2009,fang_photonic_2012,lira_electrically_2012, poulton_design_2012} & $>\Omega$ & 200 GHz - 3 THz & $>\Omega$ & 200 GHz - 3 THz\\
        Tandem modulators &\cite{doerr_optical_2011,doerr_silicon_2014,lin_compact_2019} & $\le\Omega$ & 10 GHz & $\le\Omega$ & 10 GHz \\
        \hline
    \end{tabular}
    \begin{flushleft}
        \footnotesize{%
        The parameter $\gamma$ is the total resonator linewidth, resulting from both radiative loss channels as well as waveguide coupling.
        The parameter $\Omega$ is the modulation frequency.
        The parameter $\eta$ is the modulation strength. 
        The achievable bandwidth values for the resonators assume $\Delta{\epsilon}/\epsilon = 3 \times 10^{-5}$ and $\omega/2\pi = 200$ THz.
        }
    \end{flushleft}
    \label{tab:summary_bw}
\end{table*}

In Table \ref{tab:summary_bw} we summarize both the forward bandwidth and the isolation (backward) bandwidth characteristics of all the dynamic isolator designs.
We note that for dynamic isolators the forward and backward bandwidth may not be the same, and hence here we comment on the forward and backward bandwidth separately.
For the waveguide isolators, the tandem modulator and photonic transition operate in complimentary regimes with a bandwidth that is either smaller or larger, respectively, than the modulation frequency.
Moreover, these two waveguide isolator designs have forward and backward bandwidths that are approximately equal.
In the resonant ring isolators, the isolation bandwidth is always limited by the total linewidth, $\gamma$. 
However, the three ring isolators have slightly different bandwidth constraints for the forward direction. 
The photonic transition in a ring provides broadband signal transmission in the forward direction because the signal is never affected by the modulation applied to the ring.
The Rabi splitting design has a forward bandwidth limited by the splitting between the two ring modes, which is equal to $2\eta$.
Finally, the angular momentum biased ring isolator has a forward bandwidth limited to the linedwidth $\gamma$ because under the optimal operating conditions the other mode of the ring resonator with opposite rotation is immediately adjacent to the resonance providing signal isolation.
Thus, in terms of forward signal bandwidth with low insertion loss, the photonic transition in both the ring and the waveguide provide the largest bandwidth.

\subsection{Isolation and insertion loss}

In this section we discuss the isolation and insertion loss performance of the different dynamic isolator architectures that we have reviewed.
We note that all of the designs, with the exception of the gain-loss modulated isolator, come with theoretical conditions for achieving \textit{complete} isolation and \textit{perfect} signal transmission, at least at a single operating frequency. 
However, despite such promising theoretical predictions, experimentally demonstrated dynamic isolators, especially those based on electro-optic modulation, have so far achieved only modest performance.

For example, an experimental demonstration of the indirect photonic transition [Fig. \ref{fig:waveguide_transition}(a)] in an on-chip silicon waveguide consisted of 88 individual junction diodes with alternating polarities to discretize the traveling wave modulation \cite{lira_electrically_2012}.
Although this device demonstration was extremely impressive as a proof-of-principle, in terms of performance, it only provided an isolation ratio of 3 dB and had a very high insertion loss of 70 dB in the forward direction.
The large insertion loss of the device likely had contributions both from silicon's lossy carrier injection modulation mechanism (see Sec. \ref{sec:mechanisms}), but also significant contributions from waveguide scattering losses \cite{lira_electrically_2012}, particularly from the 176 individual PN junctions that made up the waveguide core.
Overall, this experimental demonstration highlights the challenge in mitigating optical losses in the highly-complex modulation architectures required for some dynamic isolators.

For the direct photonic transition isolator [Fig. \ref{fig:waveguide_transition}(c)], there have been two experimentally demonstrated devices: one using an off-chip acousto-optic modulator \cite{li_photonic_2014a}, and one using an on-chip electro-optic modulator in silicon \cite{tzuang_nonreciprocal_2014}.
Like the on-chip experimental realization of the indirect photonic transition isolator described above \cite{lira_electrically_2012}, the direct photonic transition isolator demonstrated in \cite{tzuang_nonreciprocal_2014} is very impressive as a proof-of-principle device.
However, it also suffered from a very low isolation ratio, despite using a much simpler modulation scheme than the indirect photonic transition demonstrated in \cite{lira_electrically_2012}.
The initial proposal for the tandem isolator design (Fig. \ref{fig:waveguide_tandem}) included an on-chip experimental demonstration that achieved 11 dB of isolation and 5 dB of insertion loss \cite{doerr_optical_2011}.
While these figures are improved relative to the two photonic transition devices described above, they are still far off from the performance requirements of modern integrated photonic platforms.

For comparison, recent experimentally demonstrated magneto-optical isolators in ring resonators have achieved 20 dB of isolation with a 10 dB bandwidth of 1.6 GHz \cite{bi_onchip_2011}, while interferometric magneto-optical isolators have achieved up to 30 dB isolation with a 10 dB bandwidth of 1 THz \cite{zhang_monolithic_2019}. 
Note that the 10 dB bandwidth refers to the bandwidth over which the isolation ratio exceeds 10 dB.
The insertion loss of these magneto-optical devices was also quite large, ranging from 5 - 10 dB in interferometric devices \cite{zhang_monolithic_2019} to 19 dB in ring resonators \cite{bi_onchip_2011}.

In the remainder of this section we will focus on two of the more prominent issues that we believe are limiting the current performance of dynamics isolators: (1) fabrication variations or imperfections, such as sidewall roughness, which perturb the optical modes by an amount on the order of the modulation frequency and (2) the implementation of a sufficiently large modulation wave vector in electro-optic modulators.

\subsubsection{Fabrication variations}

Generally, fabrication variations and nonidealities, such as sidewall roughness, are always a concern in integrated photonic platforms. 
However, because the change in an optical mode frequency from these effects tends to be on the same order of magnitude as the typical modulation frequencies used in dynamic isolators. 
The effect of such structural variations on a dynamic isolator's performance may be much more significant than in other devices.
In particular, structural variations may be more of an issue for the photonic transition isolators, as well as other architectures that are designed to couple between multiple dispersion-engineered optical modes.
Considering the very long modulator lengths involved, e.g. $\sim$3.9 mm in \cite{tzuang_nonreciprocal_2014}, variations in geometric parameters such as waveguide width or height can lead to significant changes in the local dispersion of the waveguide modes. 
Thus, a modulation that is designed to be phase matched on average may not satisfy the phase matching condition at all points along the waveguide and may limit the fidelity of the nonreciprocal resposne.

Although there has not yet been an experimental demonstration of a resonant electro-optic dynamic isolator at optical frequencies, fabrication challenges such as sidewall roughness, can also be a significant issue in these devices.
The critical concern in this case comes down to how the quality factor and optical mode symmetries are affected.
Particularly in the ring resonator devices, we recall that the high rotational symmetry is essential for realizing a nonreciprocal response because it enables directional coupling between the resonator and the access waveguide \cite{yariv_coupledresonator_1999}.
Therefore it is an open question as to how backscattering may affect the non-reciprocal response in light of the high resonator quality factors that may be required.

One implementation of the angular momentum biased circulator proposed in \cite{sounas_angularmomentumbiased_2014} is to use coupled standing wave resonators that form traveling wave super modes \cite{mock_magnetfree_2019}.
However, such approaches still require high structural symmetry among their constituent resonators and can also still be highly sensitive to fabrication imperfections.
Dynamic isolator and circulator designs which do not require high structural symmetry, such as the theoretical design proposed in \cite{williamson_dualcarrier_2018}, may be more favorable for fabrication purposes.

\subsubsection{Modulation wave vector in electro-optics}

Dynamic modulation with a either a linear or angular wave vector, i.e. a traveling wave component, is a requirement for many of the isolator architectures that we have reviewed.
In principle, such a traveling component is achievable in a traveling wave electro-optic modulator, where the modulation is induced by a propagating radio frequency (RF) or microwave mode.
However, the magnitude of the wave vector required for the indirect photonic transition [Fig. \ref{fig:waveguide_transition}(a)], as well as all three ring resonator isolators [Fig. \ref{fig:resonator_transition}, Fig. \ref{fig:resonator_transition_strong}, and  Fig. \ref{fig:resonator_broken}], is very difficult to achieve in standard traveling wave electro-optic modulators.
Here, we briefly describe why this is the case and then discuss how mechanical and acoustic modulation schemes provide a compelling solution to this issue.

Standard traveling wave electro-optic modulators typically consist of transmission lines that support propagating RF modes with a wave vector on the order of $q_{\text{eo}} \sim \sqrt{\varepsilon_r(\Omega)} \frac{\Omega}{c_0}$, where $c_0$ is the speed of light and $\varepsilon_r(\Omega)$ is the average relative permittivity in the region where the modulating RF fields are concentrated \cite{johnson_high_2003}.
While such modulators do provide an ideal modulating waveform with a single spatial frequency component, the fact that the modulating frequency is far smaller than the optical signal frequency, i.e. $\Omega \ll \omega$, means that the magnitude of the spatial frequency component is much smaller than the wave vector required in dynamic isolators, i.e. $q_{\text{eo}} \ll q$ for $q$ in (\ref{eq:eps_indirect}) and (\ref{eq:eps_ring_indirect}).
Thus an open question is whether one can specifically configure the propagation characteristics of a traveling wave electro-optic modulator through concepts in metamaterial engineering.

\subsubsection{Modulation wave vector in acousto-optics and optomechanics}

In constrast to electro-optic modulation, acousto-optic modulation comes with a large \textit{built in} wave vector.
The significantly larger wave vector of acoustic modes is a direct result of the orders-of-magnitude difference between the speed of sound and the speed of light.
For a comparable modulation frequency, $\Omega$ an acousto-optic modulator provides approximately $\frac{2.99 \times 10^8 \text{ m/s}}{3.41 \times 10^2 \text{ m/s}} \approx 10^6$ larger modulation wave vector than an electro-optic modulator.
Moreover, by co-confining acoustic and optical modes in integrated waveguides, the effective strength of modulation can be enhanced significantly \cite{rakich_giant_2012}.

Compared to the relatively few experimental demonstrations of integrated nonreciprocal electro-optic devices, there have been a number of demonstrations of integrated nonreciprocal acousto-optic and optomechanical devices.
Indeed, many of these devices exploit the large wave vector available in acoustic and phononic modes and often operate analagously to several of the device architectures we have discussed in this review.
For example, there have been theoretical proposals \cite{poulton_design_2012} and experimental demonstrations \cite{kittlaus_nonreciprocal_2018a} of acoustically-driven nonreciprocal interband transitions in waveguides.
In such devices, the waveguide can be designed to simultaneously confine an optical mode and an acoustic (phonon) mode with appropriate symmetries.
Here, the phonon modal amplitude distribution plays the same role as the modulation profile in the electro-optic devices described above, meaning that an odd phonon mode is required to couple between an even and odd optical mode.
Experimental demonstrations of these devices are extremely promising from a performance point of view.
For example, the indirect transition demonstrated in \cite{kittlaus_nonreciprocal_2018a} achieved a peak isolation of 38 dB and an isolation of at least 19 dB over a broad 150 GHz bandwidth.
Unlike discretized electro-optic modulators, acoustic modulators can also provide tunable modulation wave vectors via the acoustic mode dispersion which allows for a very large change in wave vector for relatively small change in the modulation frequency.
One disadvantage of waveguide-based acousto-optic modulators is their long length where, for example, the device demonstrated in \cite{kittlaus_nonreciprocal_2018a} consisted of a 2.4 cm long waveguide.

A number of resonant acoustically-driven devices have also been theoretically proposed and experimentally demonstrated \cite{hafezi_optomechanically_2012, miri_optical_2017, ruesink_nonreciprocity_2016a, kittlaus_nonreciprocal_2018a, kim_nonreciprocal_2015, sohn_timereversal_2018, sarabalis_optomechanical_2018}.
Unlike electro-optical modulators, acousto-optic modulation schemes can be driven either optically by beating a higher power optical pump wave with a lower power detuned probe wave \cite{kittlaus_nonreciprocal_2018a, ruesink_nonreciprocity_2016a}, or electrically via surface wave or other forms of transducers \cite{sohn_timereversal_2018, vanlaer_electrical_2018}.
In terms of bandwidth, the schemes that do not rely on coupling via mechanical resonances with kHz- to MHz-scale linewidths are the most promising.

\section{Conclusion and Outlook}
\label{sec:conclusion}

In conclusion, we have reviewed recent theoretical and experimental progress on dynamically modulated optical isolators and circulators and discussed the operating principles of a number of different device architectures.
We have also analyzed performance tradeoffs between these different device architectures and highlighted a number of promising conventional and emerging modulation mechanisms that can be leveraged for constructing dynamic isolators.
In general, while there have been a number of promising theoretical proposals for dynamic isolators based on electro-optic modulation, experimental progress has achieved only modest performance.

In this review we have highlighted several of the practical challenges in this area that must be overcome in order to achieve isolator performance that can meet the demands of modern integrated photonic platforms.
One particularly important area of focus for future research is in achieving strong modulation at high speed.
Along this direction, recent efforts to integrate LiNbO$_3$ and other Pockels effect modulators are  promising.
Recent demonstrations of 2D material electro-optic modulators are also impressive for their very strong modulation, however, future work will need to focus on enhancing the effect of modulation in 2D materials on the optical mode frequency.
Generally, in electro-optic modulators, dynamic isolators will benefit strongly from achieving low-loss integration of metallic components with optical waveguides and resonators.
Finally, acousto-optic and optomechanical modulation appear poised to be a very promising platform for dynamic isolators because they provide very large built-in modulation wave vectors that break reciprocity.
However, future work on these devices, particularly those in waveguides, may focus on improving the modulation efficiency in order to reduce the length of existing centimeter-scale devices.

There is a strong motivation for dynamic isolators, particularly in application areas that are incompatible with magnetic devices. Optical sensors based on atomic transitions are one particular example of such an application space. However, other more conventional use cases, such as optical communications and laser cavity protection, can also benefit from the improved performance promised by dynamic isolators.

\section*{Acknowledgements}

The authors would like to acknowledge Prof. Meir Orenstein, Prof. Andrea Al\`{u}, Dr. Rapha\"{e}l Van Laer, Dr. Aseema Mohanty, and Dr. Viktar Asadchy for helpful discussions.

\bibliographystyle{IEEEtran}

\end{document}